\documentclass[nofootinbib,showpacs,notitlepage,prd,aps,longbibliography,onecolumn]{revtex4-1}
\usepackage{graphicx}\usepackage{amsmath}\usepackage{amssymb}\usepackage{slashed}
\usepackage{epsfig}

\newcommand{\be}{\begin{equation}}\newcommand{\ee}{\end{equation}}
\newcommand{\bea}{\begin{eqnarray}}\newcommand{\eea}{\end{eqnarray}}
\newcommand{\nn}{\nonumber}
\newcommand{\pa}{\partial}
\newcommand{\la}{\lambda}
\newcommand{\Ga}{\Gamma}
\newcommand{\ep}{\varepsilon}
\renewcommand{\phi}{\varphi}
\newcommand{\p}{\mathbf{p}}
\newcommand{\q}{\mathbf{q}}
\renewcommand{\k}{\mathbf{k}}
\newcommand{\x}{\mathbf{x}}
\newcommand{\Ref}[1]{(\ref{#1})}

\renewcommand{\v}{v_{\rm F}}
\newcommand{\elm}{electromagnetic~}
\renewcommand{\kappa}{\varkappa}
\newcommand{\om}{\omega}\newcommand{\al}{\alpha}

\newcommand{\II}{\cite{bord09-80-245406}}
\newcommand{\III}{\cite{fial11-84-035446}}

\begin{document}
\title{Surface plasmons for doped graphene}
\author{M. Bordag\footnote{bordag@itp.uni-leipzig.de}}
\affiliation{ Leipzig University, Institute for Theoretical Physics,  04109 Leipzig, Germany}
\author{I.G. Pirozhenko\footnote{pirozhen@theor.jinr.ru}}
\affiliation{ Bogoliubov Laboratory of Theoretical Physics, Joint Institute for Nuclear Research and
Dubna International University, Dubna 141980, Russia}

\date{\small \today}
\begin{abstract}
Within the Dirac model for the electronic excitations of graphene, we calculate the full polarization tensor with finite mass and chemical potential. It has, besides the (00)-component, a second form factor, which must be accounted for. We obtain explicit formulas for both form factors and for the reflection coefficients. Using these, we discuss the regions in the momentum-frequency plane where plasmons may exist and give numeric solutions for the plasmon dispersion relations. It turns out that plasmons exist for both, TE and TM polarizations over the whole range of the ratio of mass to chemical potential, except for zero chemical potential, where only a TE plasmon exists.
\end{abstract}
%
%
\maketitle
\section{Introduction}
At present, graphene is still an object of highly actual interest, especially for its optical properties and plasmonics. There is an enormous number of papers on, for recent reviews see \cite{blud13-27-1341001} and \cite{abal14-1-135}. One of the approaches to the theoretical description of its electronic excitations rests on the Random Phase Approximation (RPA) for the density-density correlation function and for the conductivity. Usually, these quantities are calculated in some approximation, the non-dispersive limit for example. Another approximation rests on the smallness of the Fermi velocity, $\v$, which is 300 times smaller than the speed of light. As shown below, this smallness allows formally to neglect one of the two form factors entering the polarization tensor.

The electronic properties of graphene are equally well described by the Dirac model, consisting of a relativistic spinor in a (2+1)-dimensional space-time (the plane of graphene). It has its own Lorentz group with $\v$ in place of $c$. The coupling to the (3+1)-dimensional electromagnetic field is the usual one. In this way, one has a modified Quantum Electrodynamics (QED), to which the well-known formalisms can be applied. For instance, one calculates the photon polarization tensor $\Pi^{\mu\nu}$. In the language of Quantum Field Theory (QFT) it consists of a fermion loop (if restricting to one loop approximation). This has been done in quite a number of papers, for instance in \cite{gusy07-21-4611} and related papers (mainly in application to external magnetic field). For a kind of review we refer to \cite{fial12-27-1260007}.  In fact, this approach is in principle equivalent to RPA, where, however, one makes some approximations for the very beginning.
The formulas obtained using the polarization tensor allow to calculate the Casimir force between graphene and, say a conducting wall, ideal \cite{bord09-80-245406} or real \cite{bord12-86-165429}, and to investigate surface plasmons, e.g. in the TE mode \cite{bord14-89-035421}. Surface plasmons were considered earlier using expressions for the conductivity obtained in RPA. For instance, in \cite{mikh07-99-016803}, using a high frequency approximation for the conductivity, a TE mode plasmon was predicted.

In the present paper, the approach from QFT is taken. The complete polarization tensor is calculated for both, mass $m$ (gap parameter) and chemical potential $\mu$, non-zero. In this case, like with finite temperature, the polarization tensor, which is transverse for gauge invariance,  has two form factors, and both are accounted for. Also, the complete frequency and momentum dependence is kept. All approximations, usually done within the RPA approach can be obtained afterwards as special cases.

The polarization tensor was previously calculated in a number of special cases. First of all one needs to mention  that the calculations in the RPA approach deliver   special cases for components of the polarization tensor. So in \cite{wuns06-8-318,hwan07-75-205418}, in fact the (00)-component of the polarization tensor, which is responsible for static screening and for the TE mode plasmon, was  calculated in the massless case. In the QFT approach, this component was calculated with chemical potential and mass in \cite{pyat09-21-025506}.

In application to graphene, the second form factor was calculated in \cite{fial11-84-035446} with mass, temperature and chemical potential, however restricted to Matsubara values of its argument, i.e., to discrete imaginary frequencies. It was applied to the calculation of the Casimir effect. Recently, it was calculated for finite temperature and mass for all, including non-Matsubara and real, frequencies \cite{bord15-91-045037}, but without chemical potential. From these papers, a representation with mass and chemical potential for real frequencies, as it is needed for plasmons, cannot be derived in any simple way. Here we fill this gap and calculate the complete polarization tensor for both non-zero, mass and chemical potential, for real frequencies.

We apply this polarization tensor to the investigation of surface plasmons. For this we start from the generic formulas for the electromagnetic field in (3+1)-dimensions and insert the polarization tensor resulting from the (2+1)-dimensional fermions. We show that this polarization tensor, which involves the Fermi speed $\v$, can be expressed in terms of the same polarization tensor calculated with unit speed. The actual calculation is done in the Appendix. Further we (re-)derive the equations for the reflection coefficients in terms of the two form factors, Eq. \Ref{1.27}, and, which is equivalent, in terms of $\Pi_{00}$ and $\Pi_{\rm tr}$, Eq. \Ref{2.6}. Further, we solve the equations for the plasmon dispersions numerically and represent the results graphically for the TE and TM-modes.

Throughout the paper we use units with $\al=e^2/(4\pi)$ for the coupling and $v=v_F/c$ for the Fermi speed and refer to
\be \al\sim\frac{1}{137}, \ v\sim\frac{1}{300},
\label{0.1}\ee
as the physical values of the parameters.

\section{Electrodynamics with polarization tensor from graphene}
By the Dirac equation model, the interaction of the electromagnetic field with the long wavelength electronic excitations in graphene can be described by relativistic quantum electrodynamics. In this formalism, the Dirac equation is
\be (i  {\slashed \pa}-e\slashed A -m)\psi=0,
\label{1.1}\ee
where
\bea &&\slashed \pa=\tilde{\gamma}^\mu\frac{\pa}{\pa x^\mu},\quad \slashed A=\tilde{\gamma}^\mu A_\mu,
\quad \tilde{\gamma}^\mu=\eta^\mu_{\mu'}\gamma^{\mu'},
\nn\\[4pt]&&\eta^\mu_{\mu'}={\rm diag}(1,v,v),
\label{1.2}\eea
with the usual gamma matrices $\gamma^{\mu}$ and the Fermi speed $v$. The resulting polarization tensor is
\be     \tilde{\Pi}^{\mu\nu}(p)=
        ie^2\int\frac{dq_0d^2q}{(2\pi)^3}\, tr \frac{1}{i\slashed q-m}\tilde{\gamma}^\mu \frac{1}{i\slashed q-i\slashed p-m}\tilde{\gamma}^\nu ,
\label{1.3}\ee
where we used momentum representation. Assuming the graphene sheet located at $z=0$, the momenta are in the directions $\mu=0,1,2$. It can be reduced to the polarization tensor $\Pi^{\mu\nu}(p)$, defined by the same formula but with $v=1$, by the substitution $q_i\to q_i/v$ ($i=1,2$) in the integration,
\be      \tilde{\Pi}^{\mu\nu}(p)=
\frac{1}{v^2}\,\eta^\mu_{\mu'}\Pi^{\mu'\nu'}(\tilde{p})\,\eta^\nu_{\nu'},
\label{1.4}\ee
where the vector $p$ is substituted by $\tilde{p}$. This vector is defined by
\be
        \tilde{p}^\mu=\eta^\mu_{\mu'} p^{\mu'}.
\label{1.5}\ee
We will use the notation with a tilde for vectors with the Fermi speed $v$ in the spatial components and for the polarization tensor \Ref{1.3} throughout the paper.

In the four dimensional formulation of electrodynamics, the Maxwell equations with polarization tensor (which we denote by a 'hat'), read
\be \pa_\mu F^{\mu\nu}+\hat{\Pi}^{\nu\mu}A_\mu=0.
\label{1.6}\ee
These can be viewed as effective Maxwell equations, which appear, e.g., integrating out the spinor fields in a functional integral representation.
Rewritten in terms  of induced current,
\be     j^\mu=\frac{-c}{4\pi}\hat{\Pi}^{\mu\nu}A_\nu,
\label{1.7}\ee
these equations are $\pa_\mu F^{\mu\nu}=(4\pi/c)j^\nu$.

Switching to 3-dimensional notations with $\mu=(0,k)$ ($k=1,2,3$),
\be     j^\mu=(c\rho,\vec{j}),\quad A^\mu=(\Phi,\vec{A}),
        \quad \vec{E}=-\vec{\nabla} \Phi-\pa_0\vec{A},
\label{1.8}\ee
we get from the transversality of the polarization tensor
\be     \hat{\Pi}^{\mu\nu}A_\mu=\hat{\Pi}^{\nu k}\pa_0^{-1}E_k,
\label{1.9}\ee
and the induced charge density and current,
\bea        \rho &=& \frac{-1}{4\pi}\,\hat{\Pi}^{0 k}\pa_0^{-1}E_k,
\nn\\[4pt]  j_k &=& \frac{-c}{4\pi}\,\hat{\Pi}^{k l}\pa_0^{-1}E_{l}.
\label{1.10}\eea
We would like to mention that from this formula the conductivity tensor can be defined as
$\sigma_{kl}=(-c/4\pi)\hat{\Pi}^{kl}\pa_0^{-1}$.
The Maxwell equations \Ref{1.4} read now
\bea        {\rm div} \vec{E} &=& 4\pi\rho,
\nn\\[4pt]  \left(-\pa_0^2+\Delta-\nabla\circ \nabla \right)\vec{E} &=& \frac{4\pi}{c}\pa_0\vec{j},
\label{1.11}\eea
or, substituting Gauss's law,
\be      \left(-\pa_0^2+\Delta \right)\vec{E} =
            \frac{4\pi}{c}\left(\nabla c\rho+\pa_0\vec{j}\right).
\label{1.12}\ee
Inserting from \Ref{1.10}, the right hand side can be expressed in terms of the polarization tensor,
\be      \left(-\pa_0^2+\Delta \right)E_k =
        -\left(\nabla_k\hat{\Pi}^{0 l}\pa_0^{-1}+\hat{\Pi}^{kl}\right)E_l.
\label{1.13}\ee
These are general formulas. For graphene we have to insert
\be     \hat{\Pi}^{\mu\nu}=\left\{
    {   \delta(z)\tilde{{\Pi}}^{\mu\nu}(p)\quad{\rm for} \ \mu,\nu=0,1,2,
    \atop   0 ~~~~~~~~{\rm for} \ \mu=3 \ {\rm or} \ \nu=3,
    }   \right.
\label{1.14}\ee
with $\tilde{\Pi}^{\mu\nu}(p)$ from \Ref{1.3}.

In momentum representation we assume all relevant quantities $\sim\exp\left(-i\om t+i\k \x \right)$
and define the vectors in the plane of graphene by
\be   \k=\left(k_1\atop k_2\right),\quad \x=\left(x_1\atop x_2\right),
\label{1.15}\ee
and similar for other vectors. Further we need to define the vector
\be\label{1.15a} p^\mu=(\om,\k,0) ,\quad p=\sqrt{\om^2-\k^2}.
\ee
Now we split equations \Ref{1.12} into two with $n,m=1,2$, i.e.,  parallel to the plane of graphene, and and a third in the perpendicular direction,
\bea        \left(p^2+\pa_z^2\right) E_n(p,z) &=& \delta(z) \Xi^{nm}E_{m}(p,0),
\nn\\[4pt]  \left(p^2+\pa_z^2\right) E_3(p,z) &=& \pa_z \delta(z)\frac{1}{i\om}\tilde{\Pi}^{0n}(p)E_n(p,0),
\label{1.16}\eea
where we introduced the notation
\be     \Xi^{nm}=\frac{p_n}{\om}\tilde{\Pi}^{0m}(p)-\tilde{\Pi}^{nm}(p),
\label{1.17}\ee
appearing in  the right side in \Ref{1.13}. In \Ref{1.16}, we have to solve the equations for the components $E_n(p,z)$ of the electric field parallel to the plane, whereas  $E_3(p,z)$ follows from $E_n(p,z)$ by integration.

Now we need a more specific expression for the polarization tensor for graphene.
From \Ref{1.4} we get
\bea\label{1.17a}   \tilde{\Pi}^{00}(p)&=&\frac{1}{v^2}\Pi^{00}(\tilde{p}),
\nn\\           \tilde{\Pi}_{\rm tr}&=&\frac{1-v^2}{v^2} \Pi^{00}(\tilde{p})
                                                        +\Pi_{\rm tr}(\tilde{p}),
\eea
where we defined
\be\label{1.17b} \Pi_{\rm tr}(\tilde{p})=g_{\mu\nu}\Pi^{\mu\nu}(\tilde{p}),\quad
                    \tilde{\Pi}(p)_{\rm tr}=g_{\mu\nu}\tilde{\Pi}(p)
\ee
for the traces.

The polarization tensor is calculated in the Appendix for $v=1$. There it is represented in terms of form factors $A(p)$ and $B(p)$, Eqs.  \Ref{A.6}, where we have to insert with \Ref{1.5}
\be \tilde{p}^\mu=(\om,v\k,0),\quad \tilde{p}=\sqrt{\om^2-(v\k)^2}
\label{1.18}\ee
in the form factors and in the tensor structures, \Ref{A.7}. For these we note
\be\begin{array}{rclrcl}
     \tilde{P}^{0m}(\tilde{p})&=& -\frac{\om k_m}{\tilde{p}^2}, &
     \tilde{P}^{nm}(\tilde{p}) & =&  -\delta_{nm}-\frac{v^2k_n k_m}{\tilde{p}^2},
\\[4pt] \tilde{M}^{0m}(\tilde{p}) &=& \frac{v^2 k^2 k_m}{\om\tilde{p}^2},&
        \tilde{M}^{nm}(\tilde{p})  &=& \frac{v^2 k_n k_m}{\tilde{p}^2} .   \end{array}
\label{1.19}\ee
Accounting also for the $\eta^\mu_{\mu'}$ we get from \Ref{1.4}
\bea     \tilde{\Pi}^{0m} &=& \frac{-\om k_m}{\tilde{p}^2}
        \left(A(\tilde{p})-\frac{v^2k^2}{\om^2}B(\tilde{p})\right),
\nn\\[4pt]   \tilde{\Pi}^{nm} &=& -\left(\delta_{nm}+
                \frac{v^2 k_n k_m}{\tilde{p}^2}\right)A(\tilde{p})
                +\frac{v^2 k_n k_m}{\tilde{p}^2}B(\tilde{p}).
\label{1.20}\eea
This allows to rewrite \Ref{1.17} in the form
\be      \Xi^{nm}=\left(\delta_{nm}-(1-v^2)\frac{  k_n k_m}{\tilde{p}^2}\right)A(\tilde{p})
                -\frac{v^2 p^2\,k_nk_m}{\om^2\tilde{p}^2}\,B(\tilde{p}).
\label{1.20M}\ee
The form factors $A({p})$ and $B({p})$ are calculated in the Appendix, Eqs.  \Ref{A.9} and \Ref{A.50}, \Ref{A.61}.

Next we introduce the polarizations for the electric field. We consider only the components $E_n$ with $n=1,2$,
\be     E_{n}(p,z)=\left({-k_2\atop k_1}\right)\Phi_{\rm TE}
                    +\left({k_1\atop k_2}\right)\Phi_{\rm TM},
\label{1.21}\ee
where, up to a normalization, we have $E_{\rm TE}=\Phi_{\rm TE}$ and $E_{\rm TM}=\pa_z\Phi_{\rm TM}$.
We do not need to consider here the third component, $E_3$, since it follows with \Ref{1.16} from $E_n$. Now, Eq. \Ref{1.16} is diagonal in the polarizations, i.e., the graphene does not mix these. So we get with \Ref{1.21} from \Ref{1.16}
\be         (p^2+\pa_z^2)\Phi_{\rm TX}(p,z)=\delta(z) \Xi_{\rm TX} \Phi_{\rm TX}(p,0).
\label{1.22}\ee
Here the subscript ${\rm 'TX'}$ stands for one of the polarizations and
\bea        \Xi_{\rm TE} &=& \frac{1}{k^2}\left({-k_2\atop k_1}\right)_nM^{nm}
                        \left({-k_2\atop k_1}\right)_m= A(\tilde{p}),
\nn\\[4pt]   \Xi_{\rm TM} &=& \frac{1}{k^2}\left({k_1\atop k_2}\right)_nM^{nm}
                        \left({k_1\atop k_2}\right)_m,
 \nn\\[4pt]&=&\frac{p^2}{\tilde{p}^2}\left(A(\tilde{p})
                        -v^2\frac{k^2}{\om^2}B(\tilde{p})\right).
\label{1.23}\eea
Finally, we rewrite the equation \Ref{1.22} for $z=0$ in terms of matching conditions,
\bea     \Phi_{\rm TX}(z=+0)- \Phi_{\rm TX}(z=-0) &=&0,
\nn\\[4pt]      \pa_z \Phi_{\rm TX}(z=+0)-\pa_z \Phi_{\rm TX}(z=+0)&=&\Xi_{\rm TX} \Phi_{\rm TX}(z=0),
\label{1.24}\eea
for both polarizations, i.e., the functions are continuous and their derivatives jump. In a standard scattering setup, the solutions are
\be     \Phi_{\rm TX}(z)=\left(e^{ipz}+r_{\rm TX}e^{-ipz}\right)  \Theta(-z)
            +t_{\rm TX}e^{ipz}\Theta(z)
\label{1.25}\ee
and the reflection and transmission coefficients are
\be     r_{\rm TX}=\frac{-1}{1+Q_{\rm TX}^{-1}},\quad t_{\rm TX}=\frac{1}{1+Q_{\rm TX}},
\label{1.26}\ee
with
\be     Q_{\rm TE}=\frac{-1}{2ip}{A}(\tilde{p}),\quad
        Q_{\rm TM}=\frac{-p}{2i\tilde{p}^2}\left(A(\tilde{p})
                        -v^2\frac{k^2}{\om^2}B(\tilde{p})\right).
\label{1.27}\ee
These are the final formulas for the coefficients. Similar formulas were derived earlier, e.g., Eq. (23) in \cite{fial11-84-035446}.

\section{Surface plasmons}
\subsection{General formulas}
Surface plasmons appear if the reflection and transmission coefficients \Ref{1.26} have a pole. So these are solutions of the equations
\be    1+ Q_{\rm TX}=0.
\label{2.1}\ee
We look for them in the frequency region
\be     vk < \om <   k   .
\label{2.2}\ee
The   upper bound implies a frequency below the continuous spectrum in order to get a wave function \Ref{1.25} decreasing to both sides of the graphene sheet.   In this region the momentum $p$ is imaginary and we use the notation
\be p=i\eta\equiv i\sqrt{-\om^2+k^2}
\label{2.3}\ee
and get from \Ref{1.27}
\be            Q_{\rm TE}=\frac{ 1}{2\eta}{A}(\tilde{p}),\quad
        Q_{\rm TM}=\frac{-\eta}{2\tilde{p}^2}\left(A(\tilde{p})
                        -v^2\frac{k^2}{\om^2}B(\tilde{p})\right).
\label{2.4}\ee
Here we insert from \Ref{A.9},
\bea        A(\tilde{p}) &=& \frac{\tilde{p}^2}{v^2k^2}\Pi^{00}(\tilde{p})
                                    +\Pi_{\rm tr}(\tilde{p}),
\nn\\[4pt]   B(\tilde{p}) &=& \frac{2\om^2\tilde{p}^2}{v^4k^4}\Pi^{00}(\tilde{p})
                                    +\frac{\om^2}{v^2k^2}\Pi_{\rm tr}(\tilde{p}).
\label{2.5}\eea
These form factors can be inserted into \Ref{2.4} and we get
\be\begin{array}{rclrl}
    Q_{\rm TE} &=& \frac{ 1}{2\eta}\left(\frac{\tilde{p}^2}{v^2k^2}\Pi^{00}(\tilde{p})
                    +\Pi_{\rm tr}(\tilde{p})\right)
        &=&\frac{ 1}{2\eta}\left(\frac{{p}^2}{k^2}\tilde{\Pi}^{00}({p})
                    +\tilde{\Pi}_{\rm tr}({p})\right),
\\[4pt]      Q_{\rm TM} &=& \frac{\eta}{2v^2k^2}\,\Pi^{00}(\tilde{p})
        &=& \frac{\eta}{2 k^2}  \, \tilde{\Pi}^{00}({p}),
\end{array}\label{2.6}\ee
where we also displayed the expressions in terms of the $\tilde{\Pi}^{\mu\nu}(p)$, Eq. \Ref{1.3}, using \Ref{1.17a}.
Eqs.  \Ref{2.6}  are the final formulas for the investigation of the equations \Ref{2.1}. The components of the polarization tensor $\Pi^{\mu\nu}(p)$ are given by Eqs.  \Ref{A.51a} and \Ref{A.61}. These coincide with Eq.(24) in \III, or Eq. (12) in \cite{bord15-91-045037}, where, however, different notations are used.

It is meaningful to check the case $\mu\le m$, which we considered in \cite{bord14-89-035421}. In that case we have $B=0$ for the second form factor and from \Ref{A.15} the relation
\be     \Pi_{\rm tr}(\tilde{p})=\frac{-2\tilde{p}^2}{v^2k^2}\Pi^{00}(\tilde{p})
\label{2.7}\ee
holds. From \Ref{2.5} we get
\be     A(\tilde{p})=-\frac{\tilde{p}^2}{v^2k^2}\Pi^{00}(\tilde{p}).
\label{2.8}\ee
In the notations of \cite{bord14-89-035421} the polarization tensor is expressed in terms of the function
\be     \Phi(\tilde{p})=
        \frac{2}{\tilde{p}} \left( 2m\tilde{p}
                -(\tilde{p}^2+4m^2){\rm arctanh}\frac{\tilde{p}}{2m}\right)
\label{2.9}\ee
by
\be         \Pi^{00}(\tilde{p})=-\frac{e^2}{4\pi}\ \frac{v^2k^2}{2\tilde{p}^2} \ \Phi(\tilde{p})
\label{2.10}\ee
delivering
\be     Q_{\rm TE}=\frac{\alpha}{2\eta}\ \frac12 \ \Phi(\tilde{p}),
\quad    Q_{\rm TM}=-\frac{\alpha\eta}{2\tilde{p}^2}\ \frac12 \ \Phi(\tilde{p}).
\label{2.11}\ee
This coincides with Eq. (18) in \cite{bord14-89-035421}.
\subsection{Formulas for the surface plasmon in the massless case}
Frequently, graphene is considered with zero gap widths, which seems justified given the small value of the gap widths. In the Dirac model this translates into the massless case. For $m=0$, the formulas \Ref{1.26} and \Ref{2.6} remain unchanged whereas those for the polarization tensor simplify. We collected them in the Appendix B. With the substitution $p\to\tilde{p}$ we get from \Ref{B.2} and \Ref{2.6}
\bea    Q_{\rm TE}&=&   \frac{4\al}{\eta}
    \left[\frac{\om^2\,\mu}{(vk)^2}-\frac{\tilde{p}}{4}\ \frac12\sum_{\la_1=\pm1}{\rm sign}(Q)
    \left(x\sqrt{x^2-1}+{\rm arccosh}(x)\right)\right],
\nn\\
     Q_{\rm TM}&=&   \frac{4\al \eta}{(vk)^2}\left[\mu
     -
     \frac{(vk)^2}{4\tilde{p}} \ \frac12\sum_{\la_1=\pm1}{\rm sign}(Q)
     \left(x\sqrt{x^2-1}-{\rm arccosh}(x)\right)\right].
\label{3.1}\eea%
The variables are
\be \eta=\sqrt{k^2-\om^2},\ \tilde{p}=\sqrt{\om^2-(vk)^2}, \ x=\frac{2\mu-\la_1\om}{vk}, \ Q=\tilde{p}^2+2\la_1\om\mu.
\label{3.2}\ee
As said above, surface plasmons appear   as solutions of the equation \Ref{2.1}.
Now, inserting numbers using \Ref{4.1} shows that for the TE polarizations the solutions start with $k>0$, whereas for the TM case the solutions start from $k=0$. In this case, the solution can be considered for small $k$.  A direct expansion of \Ref{3.1} in powers of $k$ gives
\be Q_{\rm TM}=
-\al\frac{\sqrt{k^2-\om^2}}{\om}\ \ln\frac{2\mu+\om}{2\mu-\om}+O(k).
\label{3.4}\ee
This expression is in agreement with Eq. (8) in \cite{mikh07-99-016803}, which turns out to be the approximation for small $k$ and $\om$. Eq. \Ref{2.1} with \Ref{3.4} can be solved for small $k$ by iteration. First we rewrite the equation in the form
\be \om=\al {\sqrt{k^2-\om^2}}\  \ln\frac{2\mu+\om}{2\mu-\om}+O(k).
\label{3.5}\ee
Next we insert $\om=0$ in the right side and get   $\om=\om_{\rm sf}(k)$ in the left side with
\be \om_{\rm sf}(k)=\al\,k+O(k^2).
\label{3.6}\ee
This solution confirms that we get small $\om_{\rm sf}(k)$ for small $k$ as anticipated in the derivation of \Ref{4.4}. It is interesting to note, that this solution, because of $v k<\om_{\rm sf}(k)< k$, \Ref{2.2}, is consistent for
\be v<\al<1
\label{3.7}\ee
only, which is fulfilled for the physical values of the parameters.
\subsection{Regions in the $(k,\om)$-plane}
As said in Eq. \Ref{2.2}, plasmons may exist in a region in the $(k,\om)$-plane below the continuous spectrum. In addition, the polarization tensor must be real.  Since we are in the Minkowskian region, it is real below the threshold at
\be \om_s=\sqrt{(v k)^2+(2m)^2},
\label{4.1}\ee
i.e., for $\om<\om_s$. In the case without chemical potential, i.e., for $\mu=0$, this is the only region where the polarization tensor is real. With chemical potential there is also above the threshold a region of realness. It is bounded from above by the condition \Ref{A.63a}, which reads
\be \om<\om_+^-
\label{4.2}\ee
with
\be \om_+^-=\mu+\sqrt{\mu^2+(vk)^2-2vk\,k_F}
\label{4.3}\ee
written in the notations used here. Combining \Ref{2.2} and \Ref{4.2}, we get
\be vk<\om<\min(k,\om_+^-),
\label{4.4}\ee
which defines the region in the $(k,\om)$-plane, where plasmons may exist.

\begin{figure}[h]
\epsfig{file=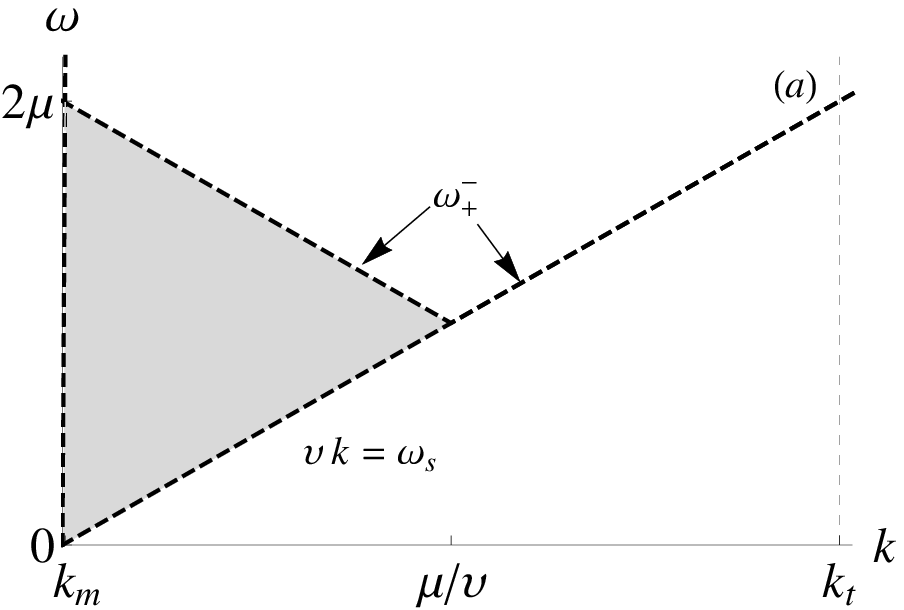,width=6cm}
\epsfig{file=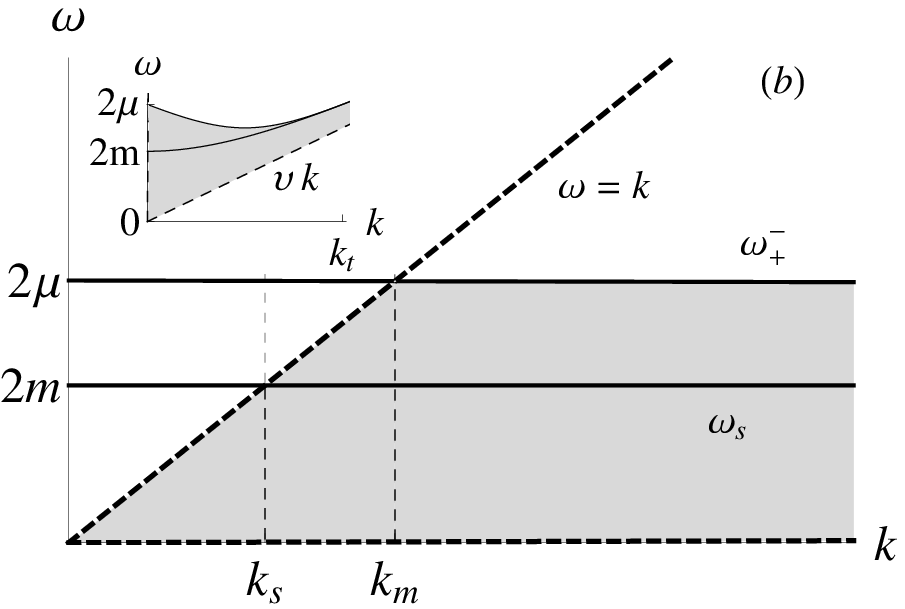,width=6cm}\\
\epsfig{file=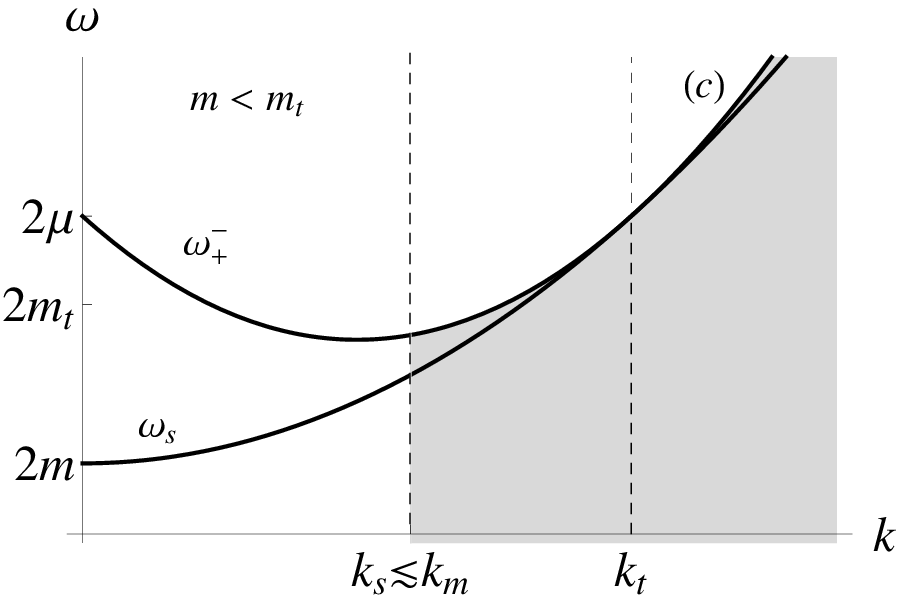,width=6cm}
\epsfig{file=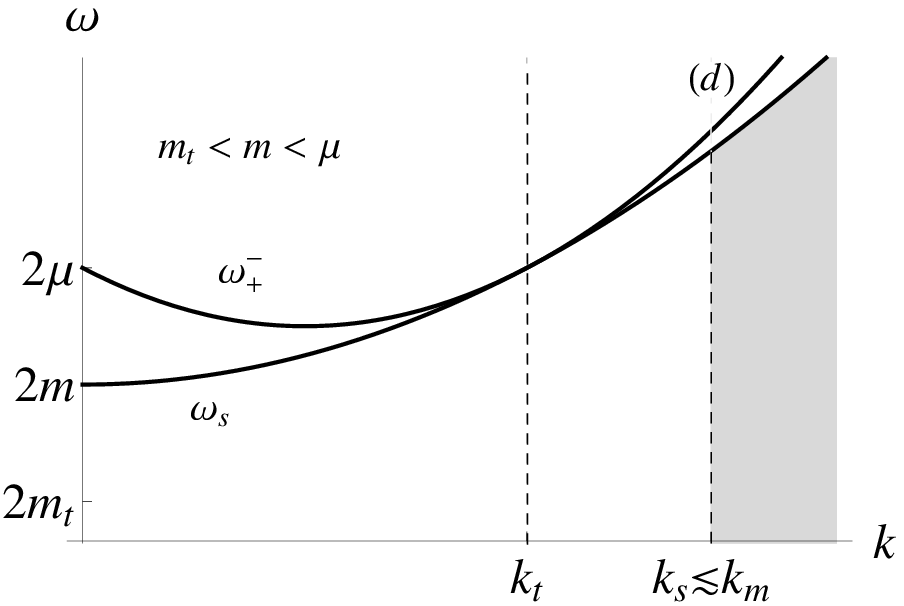,width=6cm}

\caption{The $(k,\om)$-plane with the curves for $v k$, $k$, $\om_s$ and $\om_+^-$ for  (a) $m=0$, (b) $m=0.6$, (c) $m=0.999984\mu$, i.e., $m<m_t$, and (d) $m=0.999997$, i.e., $m_t<m<\mu$, for physical values of the parameters. The region \Ref{4.4}, where a plasmon may exist, is shaded.
In panel (b), $k_t\gg k_m$ and outside the graph. The curves $\omega_p$ and $\omega_s$ appear as straight lines and the line $v k$ nearly coincides with the $k$-axis.
The inset shows the same picture for larger $k$ such that $k_t$ is seen. Now the line $\om=k$ nearly coincides with the $\omega$ axis.
Panel (d)  corresponds to the narrow region $m_t<m<\mu$, where the touching point $k_t$ is below $k_s$ and $k_m$.}
\label{regions}
\end{figure}

This region is shown in Fig.~\ref{regions}\footnote{All plots  are made for $\mu=1$.}. It looks  different for different ratios $m/\mu$.
For $m=0$ it is shown in Fig.~\ref{regions}(a). Here $\om_+^-$ degenerates, $\om_+^-=\mu+|\mu-vk|$, and the region \Ref{4.4} becomes a triangle. Since here $\om_s=vk$ holds, all solutions are automatically above the threshold.

For $m>0$ we introduce the following notations,
\be k_s=\frac{2m}{\sqrt{1-v^2}},\ k_m=\frac{2(\mu-vk_F)}{1-v^2}, \
    k_t=\frac{2k_F}{v} ,
\label{4.5}\ee
where $k_F=\sqrt{\mu^2-m^2}$ is the Fermi momentum. In  \Ref{4.5}, $k_s$ is the intersection $k=\om_s$, $k_m$ is the intersection $k=\om_+^-$, and $k_t$ is the touching point $\om_+^-=\om_s$. For  $m=m_t$, with
\be m_t=\sqrt{1-v^2}\mu,
\label{4.6}\ee
these coincide,
\be k_s=k_m=k_t=2\mu  \ \ \ \ \ (m=m_t).
\label{4.7}\ee
The case $m<m_t$ is shown in Fig. \ref{regions} (b) and $m_t<m<\mu$ is shown in Fig. \ref{regions} (d). It must be mentioned that the last region is very narrow,
\be \frac{\mu-m_t}{\mu}\sim 5\cdot 10^{-6},
\label{4.8}\ee
and $m_t=0.999994\mu$ for physical values of the parameters.
%
\subsection{TM mode surface plasmons}
TM mode surface plasmons are solutions of the equation \Ref{2.1},%
\be 1+Q_{\rm TM}=0,
\label{5.1}\ee
where $Q_{\rm TM}$ is given by \Ref{A.51a} and $\Pi^{00}(\tilde{p})$ by \Ref{A.54} with $\tilde{p}$ inserted for $p$.

\begin{figure}[h]
\epsfig{file=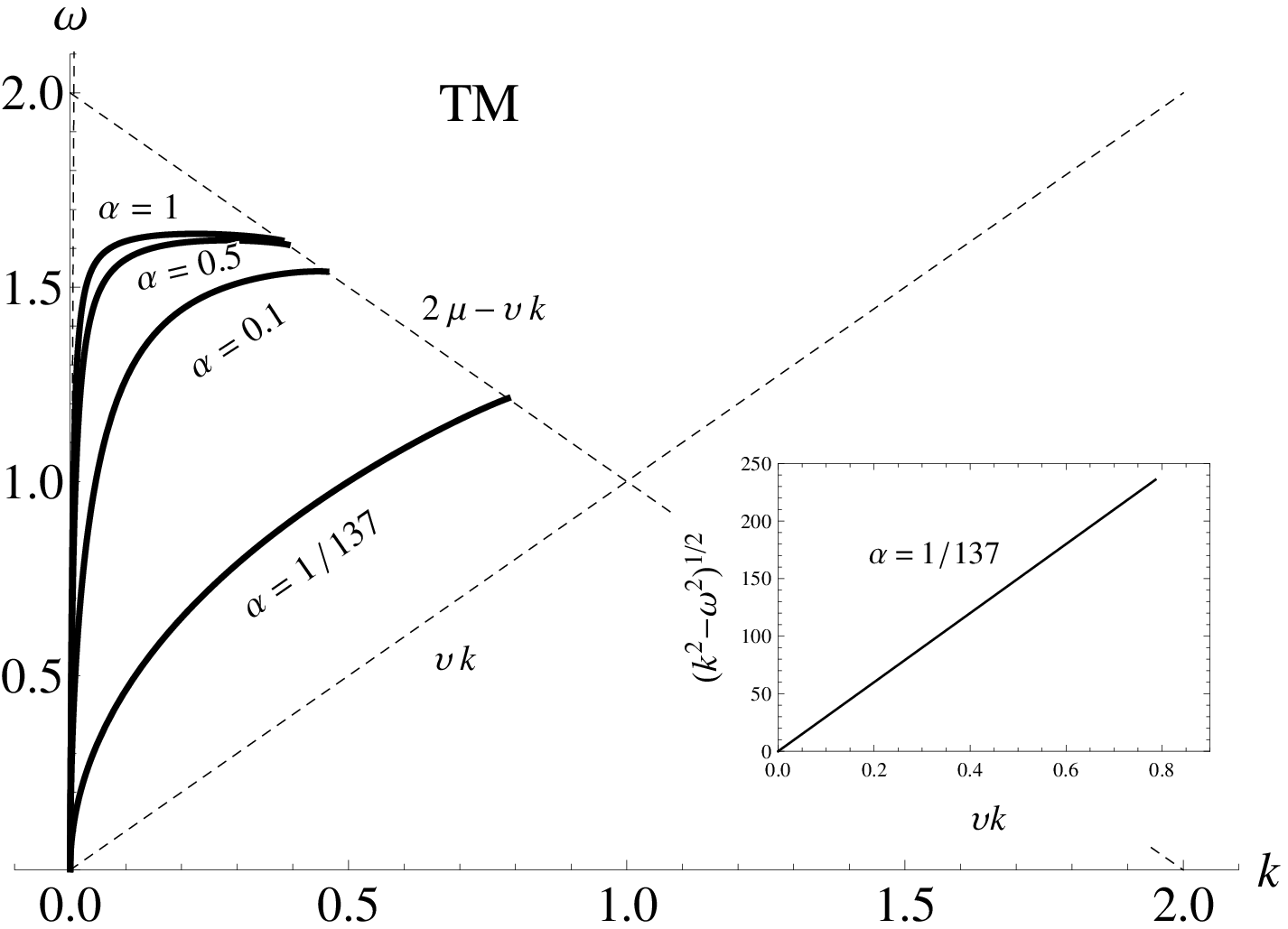,width=8cm}
\epsfig{file=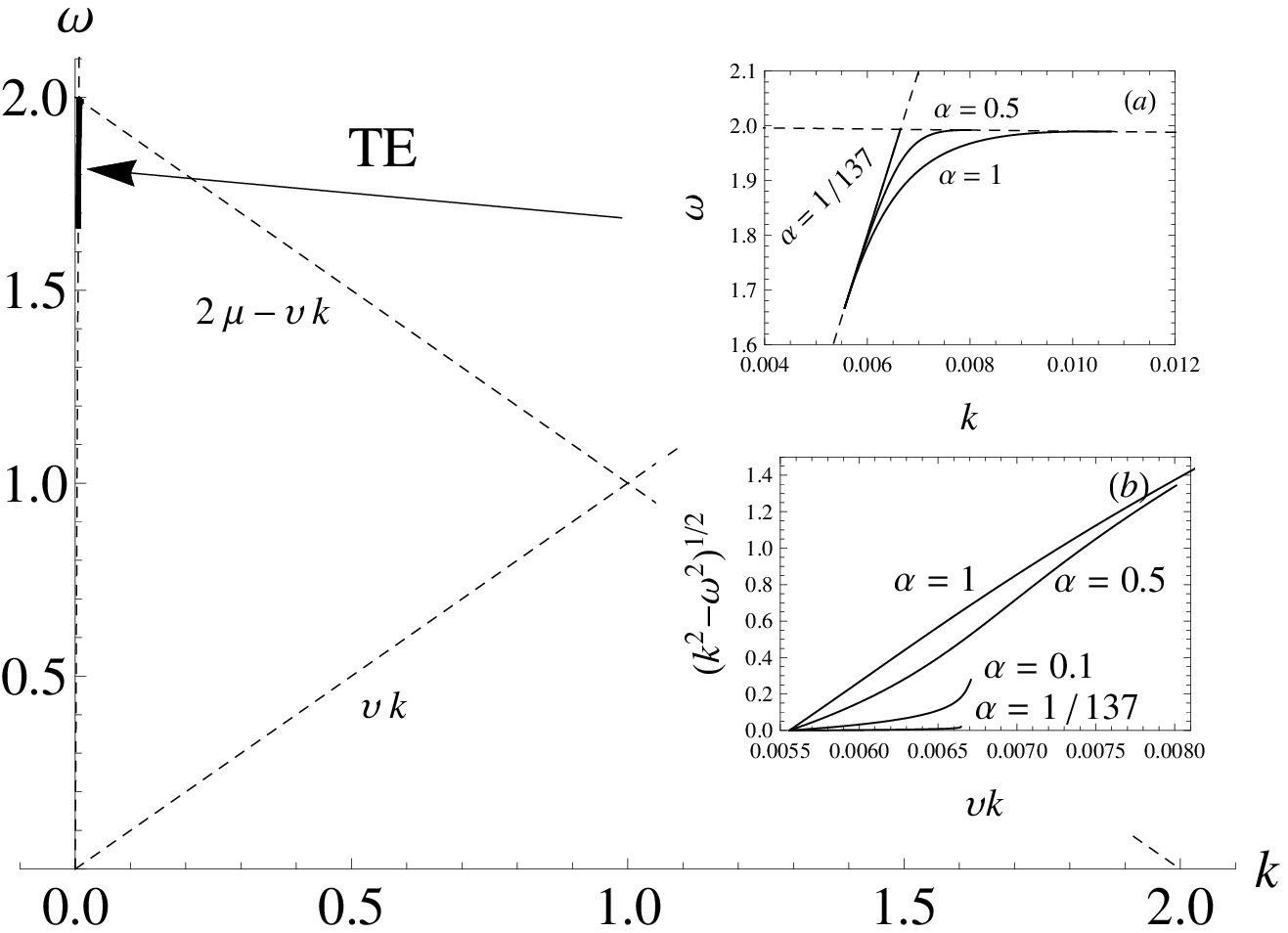,width=8cm}
\caption{Left: Transverse magnetic plasmon with $m=0$, $\mu\ne0$, and $v=1/300$,  for different values of $\alpha$. The inset shows  $\varkappa=\sqrt{k^2-\omega_{TM}^2}$ as a function
of $v k/\mu$ for $\alpha=1/137$.
 Right:   Transverse  electric plasmon. These plots are very close one to the other  differing significantly  only by their endpoints. The inset (a) is the zoom of TE curves. The nearly horizontal dashed line is  $\omega=2\mu- v k$. The inset (b) shows $\sqrt{k^2-\omega_{TE}^2}$ as a function of $ v k/\mu$.}
\label{TM1}
\end{figure}

We start with the case $m=0$. Here, $Q_{\rm TM}$ is given by \Ref{4.1}. As said above, the solution exists in the triangular region shown in Fig.~\ref{regions}(a). Examples are shown in Fig. \ref{TM1}, left panel, for several values of $\al$ and $v=1/300$. These solutions start in $k=0$ and terminates on the line $\om=2\mu-vk$.
Since these solutions are all restricted to $k<v/\mu$, the spatial extend of the solutions in the sense of $\exp(-\kappa |z|)$ is determined by $\kappa=\sqrt{k^2-\om^2}$. It is shown in Fig. \ref{TM1} in the inset.
For the physical value $\al=1/137$, the TM solution bends closer to $v k$. A similar picture may be found in \cite{stau14-26-123201}.

\begin{figure}[h]
\epsfig{file=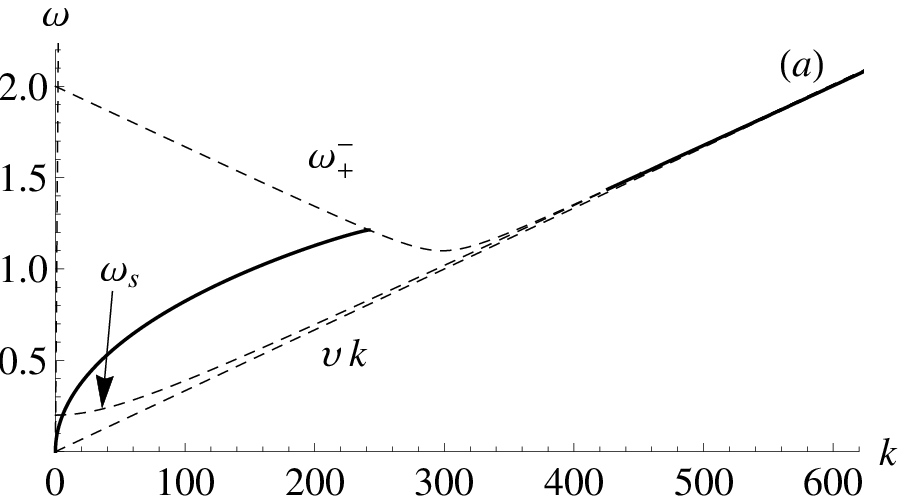,width=7cm}
\epsfig{file=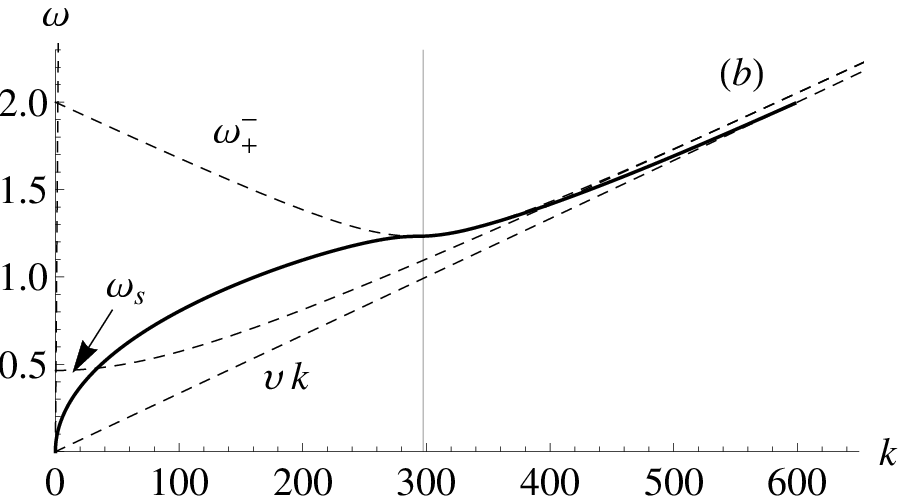,width=7cm}
\epsfig{file=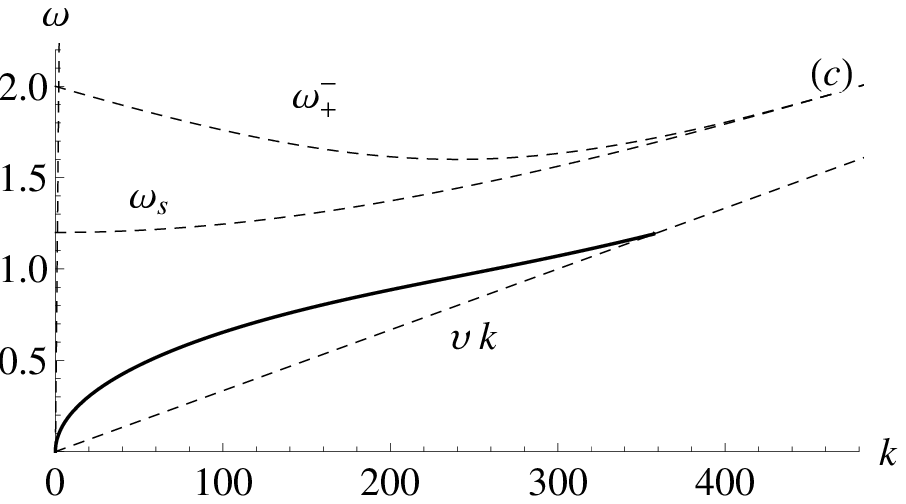,width=7cm}
\caption{TM-plasmon dispersion law for physical values of the parameters.
Panel (a) shows the solution with $m=0.1<m_0$ consisting of two branches.
Panel (b) presents the plot for $m=m_0=0.232347$, corresponding to the mass when two branches of the solution merge. For this mass  $k_{\rm max}=597.863$. Panel (c) with $m=0.6$ demonstrates that with increasing mass the endpoint moves toward the origin.
The curves for $\om_+^-$, $\om_s$ and $vk$ (from top to bottom) are shown as dashed lines.}
\label{TM2}
\end{figure}

For $m>0$, the plasmon solution exists in the regions shown in Fig.~\ref{regions}, (b)-(d). The curve $\omega_{+}^{-}$ separates from $\omega= v k$ and in the gap between them a new branch of the solution appears
for large $k$, as shown in Fig.~\ref{TM2}, panel (a).

Obviously this branch and the lower one are parts of a single solution which, however in the gap between these branches is not real. When further increasing $m$,  starting from  $m=m_0$, these branches merge, see Fig.~\ref{TM2}, panel (b).
The mass, for which the  solution touches the curve $\omega_{+}^{-}$ at its minimum, $k_{min}=\sqrt{\mu^2-m^2}/2$, is denoted by $m_0$. It can be found by substituting $k_{min}$ and $\omega_{+}^{-}(k_{min})=m+\mu$,
into Eq.~\Ref{5.1}. With physical values of the parameters,  this equation yields $m_0=0.23235\mu$.

The solution has an endpoint,
\be
k_{max}=\frac{\sqrt{3} \sqrt{8 \alpha^2 \mu (\mu-m) \left(1-v^2\right)+3 m^2 v^2}}{2 \alpha v \sqrt{1-v^2}}-\frac{3 m}{2 \alpha \sqrt{1-v^2}},
\label{5.4}
\ee
which moves down to smaller $k$ when further increasing $m$, see
Fig.~\ref{TM2}, panel (c). Finally, for $m\to\mu$, i.e. when the chemical potential disappears, this endpoint goes down to $k=0$, and the solution disappears.

For small $k$ it is possible to find the solution explicitly. For this, we rewrite Eq.~\Ref{5.1} in the form
\be \omega=\sqrt{k^2-4 v^2 k^4 \Pi_{00}^{-2}},
\label{5.4a}\ee
which allows for iteration with inserting  $\omega=k$ in the right side,
\be \omega=k -\frac{v^4}{8 \alpha^2 \mu^2}
    \frac{1}{(1 - \frac{\mu}{\sqrt{m^2 v^2 + \mu^2 (1 - v^2)}})^2}\,k^3+O(k^4).
\label{5.5}\ee
The  coefficient in front of  $k^3$ is negative, showing that the solution goes indeed below $\omega=k$, i.e., below the
border of the continuous spectrum.
This coefficient is small, proportional to $(v^2/\al)^2\sim 2.3\cdot10^{-6}$, unless $m\to\mu$, where it becomes infinite and the expansion breaks down (expansion \Ref{3.6} holds instead).
%
\subsection{TE mode surface plasmon}
\begin{figure}[h]
\epsfig{file=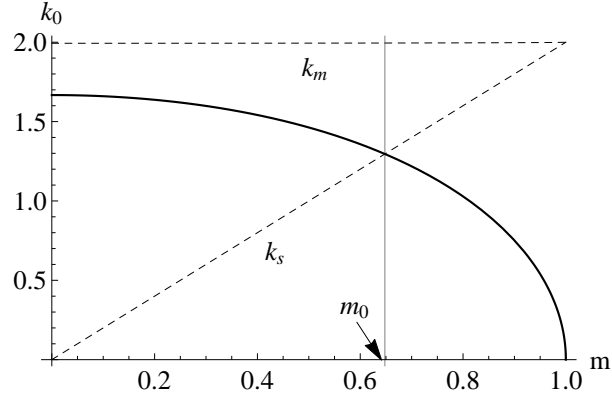,width=8cm}
\caption{The starting  point $k_0$ of the TE-plasmon as a function of $m$ for physical values of $\al$ and $v$, together with $k_s$ and $k_m$.}
\label{TE1}
\end{figure}

TE mode surface plasmons are solutions of the equation \Ref{2.1},
\be 1+Q_{\rm TE}=0,
\label{6.1}\ee
where $Q_{\rm TE}$ is given by \Ref{2.6} with \Ref{A.51a} and \Ref{A.61} with $\tilde{p}$ inserted for $p$.
The TE mode surface plasmon solution has a nonzero staring point, which we denote by $k_0$. It can be found as a solution of the equation $1+{Q_{\rm TE}}_{|_{\om=k}}=0$. For $k<k_0$, the solution goes into the continuous spectrum.  It is shown in Fig. \ref{TE1}, together with $k_s$ and $k_m$, Eq. \Ref{4.5}, as a function of $m$. The intersection of these curves in the left panel is denoted by $m_0$. This means, for $m>m_0$, that the starting point  in the $(k,\om)$-plane, see Fig. \ref{regions}, is to the {left} of $k_s$ and for $m<m_0$ to the {right}. The latter implies that the starting point is {above} the threshold.

In the case $m=0$, as mentioned above, the solution exists in the triangular region shown in Fig. \ref{regions}, panel (a). Examples are shown in Fig.~\ref{TM1}, right panel, for several values of $\al$ and $v=1/300$.
The TE solution goes close to $\omega=k$. Its frequency lies in the band $1.667<\omega/\mu<1.9934$. This is in agreement with~\cite{mikh07-99-016803}. The starting point lies at $\omega=k$ and does not depend on $\alpha$. The endpoint is situated on the line $\omega=2\mu- v k$ and depends on $\alpha$, see the inset. With decreasing $\alpha$ it tends to $k_{m}=2\mu/(1+v)$.

\begin{figure}[h]
\epsfig{file=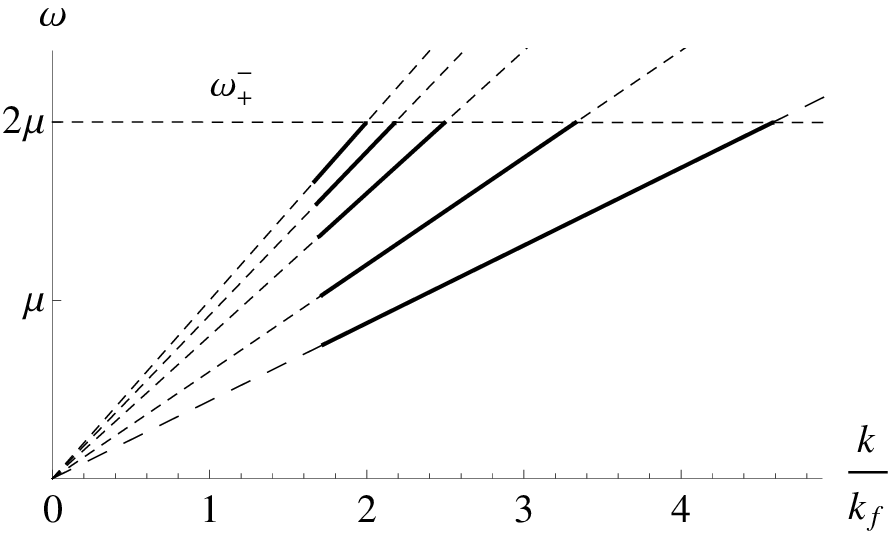,width=6cm}
\epsfig{file=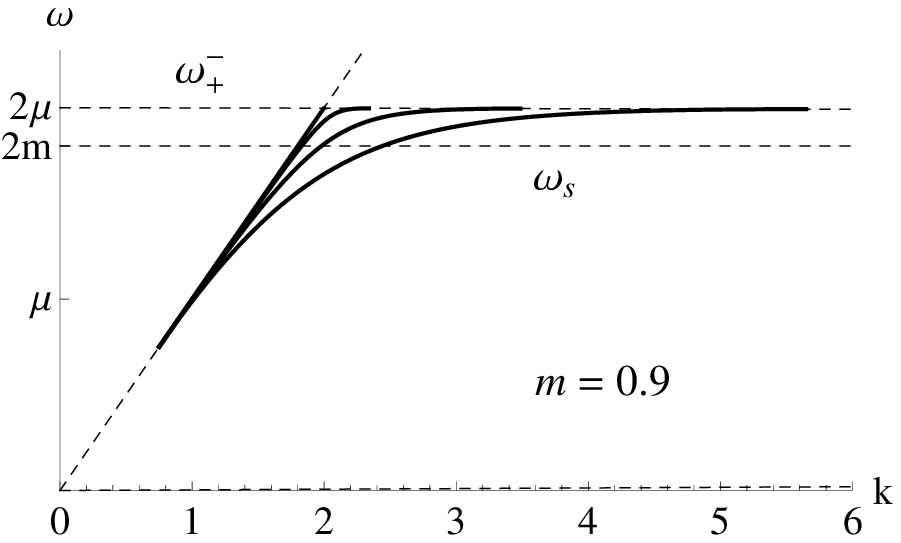,width=6cm}
\caption{Left panel: TE-plasmon dispersion laws for physical values of the parameters (solid lines) for several values of the mass, $m=0, 0.4, 0.6, 0.8, 0.9$, as function of $k/k_F$. The uppermost curve   corresponds to zero mass and is the same as in Fig. \ref{TM1}, right panel.
Right panel: the same for $m=0.9$, for  several values of the coupling, $\alpha=1, 0.5, 0.2, 1/137$. The knee at the end of the curves in the left panel becomes visible only for  coupling stronger than its physical value. The curves for $\om_+^-$ and $\om_s$ are shown as dashed lines. These are nearly horizontal. Also the straight lines for $\om=k$ are shown as dashed lines. The corresponding line for $vk$ coincides in the plot with the $k$-axis.}
\label{TE2}
\end{figure}

For $m<m_t$, the solution ends on the line $\om_+^-$, thus above the threshold. Such solutions are shown in Fig. \ref{TE2} for several $m<m_t$. In the $(k,\om)$-plane, all appear nearly on one and the same line. Therefore we represented them in the $(k,\om/k_F)$-plane. Close to the end, these lines have a knee. This is shown in Fig. \ref{TE2} in the right panel for a particular value of $m$ using smaller values of $\al$ enlarging this part.

Further increasing $m$, for $m>m_t$, the endpoint appears below the threshold and it goes to larger $k$ the closer $m$ comes to $\mu$. At the same time the starting point goes to zero, see Fig. \ref{TE1}. For $m\to\mu$, which is the transition to the case without chemical potential, the starting point is zero and the endpoint goes to infinity. The picture turns into that shown on Fig. 2 in \cite{bord14-89-035421}.

\section{Conclusions}
In the foregoing sections we calculated the polarization tensor for graphene with mass and chemical potential using formalism and notations of quantum field theory.

The actual calculation of the polarization tensor is carried out in the Appendix, for unit speed of light. In the case considered here, all integrations can be carried out and explicit formulas for the complete polarization tensor emerge. It is seen that it does not have an ultraviolet divergence, which is the expected result from the interplay of gauge invariance and dimensionality.

The  polarization tensor with $m$ and $\mu$ has 2 form factors. The result is given in terms of these. Formulas of this kind for the polarization tensor were obtained previously, but never in such completeness as here. Especially, in  \cite{pyat09-21-025506}, only one form factor, $\Pi^{00}(\tilde{p})$, was calculated.

It should be mentioned, that the second form factor, $\Pi_{\rm tr}(\tilde{p})$, Eq.\Ref{A.9tr}, frequently gives only a small contribution. This can be seen from the upper line in Eq. \Ref{2.6}, where $\Pi_{\rm tr}(\tilde{p})$ enters with a relative weight proportional to $v\sim 1/300$   as compared to $\Pi^{00}(\tilde{p})$. In other cases, its contribution is essential. For example, for $\mu\le m$, from \Ref{2.7}, $\Pi_{\rm tr}(\tilde{p})$ gives twice the contribution as $\Pi^{00}(\tilde{p})$ and has opposite sign. Thus dropping $\Pi_{\rm tr}(\tilde{p})$ in $Q_{\rm TE}$, Eq. \Ref{2.6}, changes even the sign, which results in the disappearing of the TE plasmon in this case.

The explicit form of the final formulas looks different in different regions in the $(k,\om)$-plane. All these are related by corresponding analytic continuation, which can be best understood in terms of the left side of Eq. \Ref{A.36}, viewed as an integral representation of the result. For the regions relevant for plasmons, i.e, where the polarization tensor is real, we gave explicit expressions in terms of real functions, Eqs.  \Ref{A.51a}, \Ref{A.61} for frequencies below the threshold, and Eqs.  \Ref{A.65}, \Ref{A.66} for above. Also, we gave in Appendix B explicit formulas for the massless case.

In section II we started from the general 4-dimensional notations of QED, used the effective Maxwell equations and related these to the (3+1)-dimensional formulation with (2+1)-dimensional polarization tensor. This allowed, after  separation of the polarizations into TE and TM, to formulate a scattering setup for the \elm field and to relate the scattering coefficients with the form factors of the polarization tensor. We represented the whole derivation in necessary detail to make the paper self contained. Also, at the end of the section, we discussed the restoration of $v\ne1$, including the mixing of the from factors, Eq. \Ref{1.17a}.
Finally, in Sec. III.A, from the reflection coefficients, we derived the equations for the plasmons, including the simplifications appearing for $m=0$, Sec. III.B.

In the remaining part of Sec. III, we investigated the plasmons. There, a quite sophisticated number of cases appeared. First of all, the regions in the $(k,\om)$-plane, where plasmons exist, quite strongly depend on $m$ in the interval $0\le m\le\mu$, see Fig. \ref{regions}. In the massless case, we reproduce known results, see Fig. \ref{TM1}. When the mass starts to grow, these pictures deform, see Figs. \ref{TM2} and \ref{TE2}. Finally, for $m=\mu$, i.e., when the chemical potential disappears, we turn into the case considered in \cite{bord14-89-035421}.

A general property of these solutions is that there are always (except for $\mu=0$) solutions for both polarizations present. Thereby, for small $m$, the TM solution is present starting from $\om\sim k\ge0$ and the TE solution only for larger $k$ and $\om$, see Fig.~\ref{TM1}. For $m$ closer to $\mu$, the range of $k$, where the TM solution exists, shrinks, Fig. \ref{TM2},  panel (c), whereas the TE solution extends to larger $k$-intervals, see Fig. \ref{TE2}, right panel. Roughly speaking, for $m=0$, $\mu\ne0$, there is a TM solution and for $m\ne0$, $\mu=0$, there is a TE solution.

It must be mentioned that, possibly, not all solutions considered here, are physical in the sense that some, e.g., Fig. \ref{TM2}, left panel, exist only for $k$ exceeding the range of validity of the Dirac model for graphene. Another observation is that the smallness of the physical parameters, enters in different way. Besides the general proportionality to $\al$ of the polarization tensor, and to the Fermi speed $v$ in the range \Ref{2.2}, the combination $v^2/\al$ enters, Eq. \Ref{5.5}. The combination $\al/v$, which is of order of unity, did not play any pronounced role.

It would be interesting to carry on a similar analysis of the plasmons for finite temperature without and with chemical potential.


\acknowledgements
We acknowledge a partial support from the Heisenberg-Landau Programme.

\appendix
\section{Polarization tensor in $(2+1)$ dimensions with chemical potential}
In this appendix we display the calculation of the polarization tensor.
We consider a space-time with metric $g_{\mu\nu}={\rm diag}(1,-1,-1)$ with Greek indices $\mu,\nu=0.1.2$. For the spatial part we use Latin indices $i,j=1,2$. As said in Section II, a metric with a Fermi speed  $\v\ne1$ can be restored afterwards. Following \cite{fial09-42-442001,bord09-80-245406,fial11-84-035446}, the polarization tensor is
\be \Pi^{\mu\nu}(p)=i2e^2\int\frac{dq_0d^2q}{(2\pi)^3}\, {\rm tr}\, S(q)\gamma^\mu S(q-p)\gamma^\nu
\label{A.1}\ee
with the spinor propagator
\be S(q)=\frac{i\slashed q+m}{q^2-m^2-i0}
\label{A.1a}\ee
and for the 4-dimensional gamma matrices  $\{\gamma^\mu,\gamma^\nu\}=2g^{\mu\nu}$ holds.
The factor 2 in front comes from the number $N_f=2$ of flavors.
In this appendix vectors are denoted by
\be\begin{array}{rclrcl}
    p^\mu&=&(p_0,\bar{\mathbf{p}}),& \quad \bar{\mathbf{p}}&=&(p_1,p_2) ,\\[4pt]
        p&=&\sqrt{p_0^2-\bar{p}^2},&\quad \bar{p}&=&\sqrt{p_1^2+p_2^2}.
\end{array}\label{A.1b}\ee
In this appendix we use lower indices for the time components, $\Pi_{00}=\Pi^{00}$, for convenience.
The units are taken such that $\alpha =e^2 /4\pi$ holds. Carrying out the trace in \Ref{A.1} we get
\be \Pi^{\mu\nu}(p)=i8e^2\int\frac{dq_0d^2q}{(2\pi)^3}\,\frac{Z^{\mu\nu}}{N},
\label{A.2}\ee
where
\bea Z^{\mu\nu}&=&q^\mu(q^\nu-p^\nu)+(q^\mu-p^\mu)q^\nu-q(q-p)g^{\mu\nu}+m^2g^{\mu\nu},
\nn\\[4pt] N&=&(q^2-m^2+i0)((q-p)^2-m^2+i0),
\label{A.3}\eea
with $q^\mu=(q_0,\q)$, $\bar{q}=|\q|$.
For instance, we note
\be Z_{00}= q_0(q_0-p_0)+ {\q}({\q}-{\p}) +m^2.
\label{A.4}\ee
The $q_0$-integration in \Ref{A.1} is specified by the causal $'\!\!+i0'$-description. The polarization tensor \Ref{A.1} is transversal,
\be p_\mu \Pi^{\mu\nu}(p)=0.
\label{A.4a}\ee
Before carrying out the integration this manifests itself in the structure of the numerator which can be written as
\be p_\mu Z^{\mu\nu}=-((q-p)^2-m^2)q^\nu+(q^2-m^2)(q-p)_\nu.
\label{A.5}\ee
In each term a factor from the denominator \Ref{A.3} cancels and the remaining integrals correspond to  tadpole diagrams and vanish due to parity.

In case of chemical potential or temperature, the polarization tensor depends, besides on $p^\mu$, also on an additional vector $n^\mu=(1,0,0)$. From transversality we have 2 independent tensor structures,
\be \Pi^{\mu\nu}({p})=P^{\mu\nu}(p) A(p)+M^{\mu\nu}(p) B(p),
\label{A.6}\ee
with
\be P^{\mu\nu}(p)=g^{\mu\nu}-\frac{p^\mu p^\nu}{p^2}, \quad
    M^{\mu\nu}(p)=\frac{p^\mu p^\nu}{p^2}-\frac{p^\mu n^\nu+n^\mu p^\nu}{np}
    +n_\mu n_\nu\frac{p^2}{(np)^2}.
\label{A.7}\ee
Below we will need the special cases
\be \begin{array}{rclrcl}
    P_{00}(p)&=&-\frac{\bar{p}^2}{p^2},&\quad M_{00}(p)&=&\frac{\bar{p}^4}{p_0^2p^2} , \\[4pt]
    g_{\mu\nu}P^{\mu\nu}(p)&=&2,&
            g_{\mu\nu}M^{\mu\nu}(p)&=&-\frac{\bar{p}^2}{p_0^2}.
\end{array} \label{A.8}\ee
The form factors can be obtained from \Ref{A.6},
\bea A(p)&=&\frac{p^2}{\bar{p}^2}\Pi_{00}(p)+\Pi_{\rm tr}(p),\nn\\
    B(p)&=&2\frac{p_0^2p^2}{\bar{p}^4}\Pi_{00}(p)+\frac{p_0^2}{\bar{p}^2}\Pi_{\rm tr}(p),
\label{A.9}\eea
where
\be \Pi_{\rm tr}(p)\equiv  g_{\mu\nu}\Pi^{\mu\nu}(p)
\label{A.9tr}\ee
is the trace over the Lorentz indices. We will use this subscript also for other tensors.
Further we note
\be \Pi^{0n}=\frac{p_0p_n}{\bar{p}^2}  \Pi_{00},
\label{A.900}\ee
which is a direct consequence of the transversality \Ref{A.4a}.

For the calculation of $\Pi_{\rm tr}$ we mention with \Ref{A.3}
\bea Z_{\rm tr}&=&-q(q-p)+3m^2\nn\\&=&-\frac12(q^2-m^2+(q-p)^2-m^2)+\frac12(p^2+4m^2),
\label{A.11}\eea
where in the second line the $q$-dependent terms are written in a way cancelling the corresponding factors in the denominator, and a term, which is independent of $q$. In this way we get
\be \Pi_{\rm tr}(p)=-\Sigma_{\rm tp}+\frac{p^2+4m^2}{2}\,\Sigma(p),
\label{A.12}\ee
where
\be \Sigma_{\rm tp}=i8e^2\int\frac{dq_0d^2q}{(2\pi)^3}\,\frac{1}{q^2-m^2+i0}
\label{A.13}\ee
corresponds to a tadpole  graph and
\be \Sigma(p)=4ie^2\int\frac{dq_0d^2q}{(2\pi)^3}\,
    \frac{1}{(q^2-m^2+i0)((q-p)^2-m^2+i0)}
\label{A.14}\ee
is the scalar loop.

We mention that in case there is no vector $n^\mu$ in $P^{\mu\nu}$, the tensor structure is given by the first term in \Ref{A.6} alone and the relation
\be \Pi_{00}(p)=-\frac{\bar{p}^2}{2p^2}\Pi_{\rm tr}(p)
\label{A.15}\ee
must hold. In this way, the calculation of the full polarization tensor can be reduced to the calculation of $\Pi_{00}$ or of the scalar loop $\Sigma$ and the tadpole $\Sigma_{\rm tp}$.

Next we calculate $\Pi_{00}$ and $\Pi_{\rm tr}$ directly. In the calculation we consider $\Pi_{00}$, $\Sigma$ and  $\Sigma_{\rm tp}$ in parallel since most steps in the calculation are the same for them.
Starting from here we include the chemical potential $\mu$. For this we include for a moment temperature too using the Matsubara representation. Thus we substitute $p_0$ by the Euclidean momentum, $p_0=ip_4$ with $p_4=2\pi l T$ (l integer) and we turn the $q_0$-integration to the imaginary axis and include the chemical potential, $q_0=iq_4-\mu$ with $q_4=2\pi(n+1/2)T$ (n-integer). Indicating $\mu$ in the argument we get
\bea    \Pi_{00}(p;\mu)&=&-8e^2T\sum_n \int\frac{d^2q}{(2\pi)^2}
        \frac{(iq_4-\mu)(iq_4-ip_4-\mu)+{\q}({\q}-{\p})+m^2}
                {\left[(iq_4-\mu)^2-\Gamma_1^2\right]\left[(iq_4-ip_4-\mu)^2-\Gamma_2^2\right]},
                \nn\\[4pt]
        \Sigma_{\rm tp}(p;\mu)&=&-8e^2T\sum_n \int\frac{d^2q}{(2\pi)^2}
        \frac{1}{(iq_4-\mu)^2-\Gamma_1^2},
        \nn\\[4pt]
        \Sigma(p;\mu)&=&-8e^2T\sum_n \int\frac{d^2q}{(2\pi)^2}
        \frac{1} {\left[(iq_4-\mu)^2-\Gamma_1^2\right]\left[(iq_4-ip_4-\mu)^2-\Gamma_2^2\right]}
\label{A.17}\eea
with the notations
\be \Ga_1=\sqrt{{\q}^2+m^2},\quad \Ga_2=\sqrt{({\q}-{\p})^2+m^2}.
\label{A.Ga}\ee
We mention, that $\Pi_{00}$ in \Ref{A.17} coincides up to an overall factor with Eq. (A.3) in \cite{pyat09-21-025506}.

Further in this paper we restrict ourselves to $T=0$. So we go back from the Matsubara summation to the integration, $T\sum_n\to\int dq_4/(2\pi)$.
To proceed, we rewrite $ \Pi_{00}(p;\mu)$ and $\Sigma(p;\mu)$ in a form with denominators linear in $q_4$,
\bea \Pi_{00}(p;\mu)&=&-8e^2 \int\frac{dq_4 d^2q}{(2\pi)^3} \,
            \frac14\sum_{\la_1,\la_2=\pm1}
                    \frac{M_{\la_1\la_2}}{\left[iq_4-\mu+\la_1\Ga_1\right]
                                            \left[iq_4-ip_4-\mu+\la_2\Ga_2\right]},
\nn\\[4pt]
    \Sigma(p;\mu)&=&-8e^2 \int\frac{dq_4 d^2q}{(2\pi)^3} \,
            \frac14\sum_{\la_1,\la_2=\pm1}\frac{\la_1\la_2}{\Ga_1\Ga_2}
                    \frac{1}{\left[iq_4-\mu+\la_1\Ga_1\right]
                                           \left[iq_4-ip_4-\mu+\la_2\Ga_2\right]},
\label{A.22}\eea
with
\be M_{\la_1\la_2}=1+\frac{\la_1\la_2}{\Ga_1\Ga_2}({\q}({\q}-{\p})+m^2).
\label{A.21}\ee
This rewriting is done in such a way, that the momentum $q_4$ appears in the denominator only.

Now, in order to proceed, we need to handle the ultraviolet divergence. In general, there are none in the polarization tensor in $(2+1)$ dimensions. This is obvious from power counting in \Ref{A.17}. We have 3 powers from the integration and -2 powers from the propagators. Due to the gauge invariance, 2 divergent powers drop out and a convergent expression is left. However, in intermediate steps, before the compensation due to transversality is on work, there are divergences. Therefore we assume a regularization not breaking the gauge invariance and not affecting the $q_4$-integration. Obviously this is possible. An example is dimensional regularization in the spatial directions.

With such regularization assumed we carry out the $q_4$-integration in \Ref{A.22} using
\be \int_{-\infty}^\infty\frac{dx}{(ix+a)(ix+b)}=\pi\frac{{\rm sgn}(Re(a))-{\rm sgn}(Re(b))}{-a+b},
\label{A.19}\ee
which follows simply with the Cauchy theorem.  With $a\to -\mu+\la_1\Ga_1$ and $b\to -ip_4-\mu+\la_2\Ga_2$ we get
\bea \Pi_{00}(p;\mu)&=&-8e^2 \int\frac{ d^2q}{(2\pi)^2} \,
            \frac18\sum_{\la_1,\la_2=\pm1}
                    {M_{\la_1\la_2}}
                    \frac{{\rm sgn}(-\mu+\la_1\Ga_1)-{\rm sgn}(-\mu+\la_2\Ga_2) }
                    { -\la_1\Ga_1-ip_4+\la_2\Ga_2},
\nn\\[4pt]
    \Sigma(p;\mu)&=&-8e^2 \int\frac{ d^2q}{(2\pi)^2} \,
            \frac18\sum_{\la_1,\la_2=\pm1}\frac{\la_1\la_2}{\Ga_1\Ga_2}
                    \frac{{\rm sgn}(-\mu+\la_1\Ga_1)-{\rm sgn}(-\mu+\la_2\Ga_2)}{ -\la_1\Ga_1-ip_4+\la_2\Ga_2}.
\label{A.23}\eea
The signs factor in the numerators   can be rewritten as
\be {\rm sgn}(-\mu+\la_1\Ga_1)-{\rm sgn}(-\mu+\la_2\Ga_2)=
\la_1-\la_2-2\left(\la_1\Theta(\la_1\mu-\Ga_1)-\la_2\Theta(\la_2\mu-\Ga_2)\right)
\label{A.24}\ee
using the step function. Now, for $\mu\le m$, the expression in the parentheses vanishes and only the contribution without chemical potential is left. This allows to separate the contributions from the chemical potential,
\bea \Pi_{00}(p;\mu)&=&\Pi_{00}^{(0)}(p)+\Delta_\mu \Pi_{00}(p),
\nn\\[4pt]
        \Sigma(p;\mu)&=&\Sigma^{(0)}(p)+\Delta_\mu \Sigma(p),
\label{A.25}\eea
where the superindex '(0)' denotes the vacuum part and
\bea        \Delta_\mu \Pi_{00}(p)&=&-8e^2\int\frac{d^2q}{(2\pi)^2} \,
            \frac14\sum_{\la_1,\la_2=\pm1} M_{\la_1\la_2} \,\frac{ \la_1\Theta(\la_1\mu-\Ga_1)-\la_2\Theta(\la_2\mu-\Ga_2) )}{ip_4+\la_1\Ga_1-\la_2\Ga_2},
\nn\\[4pt]    \Delta_\mu \Sigma(p)&=&    -8e^2\int\frac{d^2q}{(2\pi)^2} \,
            \frac14\sum_{\la_1,\la_2=\pm1}    \frac{\la_1\la_2}{\Ga_1\Ga_2}
              \,  \frac{ \la_1\Theta(\la_1\mu-\Ga_1)-\la_2\Theta(\la_2\mu-\Ga_2) }{ip_4+\la_1\Ga_1-\la_2\Ga_2}.
\label{A.26}\eea
are the additional contributions from the chemical potential.
Similar separation takes place for the tadpole contribution which, however, will be calculated later directly.

In the above expression, the momentum integration is bounded by $\Ga_1\le|\mu|$, $\Ga_2\le|\mu|$ and the integration is finite. This is in line with the general situation that in the polarization tensor in an external field (e.g., a magnetic field) or with temperature can be separated into the vacuum part and the field or temperature dependent part, which is free of ultraviolet divergences (at least in one loop).

Once the momentum integration is bounded, we can remove any regularization and calculate the integrals directly.
For this we  exchange in the second term in the numerators $\Ga_1\leftrightarrow\Ga_2$, $\la_1\leftrightarrow\la_2$. The exchange of the $\Ga$'s appears from the substitution $\bar{q}\to \bar{p}-\bar{q}$ of the integration variable in \Ref{A.26}. We mention that this is in general not possible if a regularization, a momentum cut-off for example, is present. From \Ref{A.26} we get this way
\bea        \Delta_\mu \Pi_{00}(p)&=&-8e^2\int\frac{d^2q}{(2\pi)^2} \,
            \frac14\sum_{\la_1,\la_2=\pm1} M_{\la_1\la_2} \,\Theta(\la_1\mu-\Ga_1)
            \left[\frac{ -\la_1}{ip_4-\la_1\Ga_1+\la_2\Ga_2}
                    +\frac{ \la_1}{ip_4+\la_1\Ga_1-\la_2\Ga_2}\right],
\nn\\[4pt]    \Delta_\mu \Sigma(p)&=&    -8e^2\int\frac{d^2q}{(2\pi)^2} \,
            \frac14\sum_{\la_1,\la_2=\pm1}    \frac{\la_1\la_2}{\Ga_1\Ga_2} \,
               \Theta(\la_1\mu-\Ga_1)
            \left[\frac{ -\la_1}{ip_4-\la_1\Ga_1+\la_2\Ga_2}
                    +\frac{ \la_1}{ip_4+\la_1\Ga_1-\la_2\Ga_2}\right].
\label{A.27}\eea
Next, we reorder the summation by substituting $\la_2=\sigma\la_1$,
\bea        \Delta_\mu \Pi_{00}(p)&=&-8e^2\int\frac{d^2q}{(2\pi)^2} \,
            \frac14\sum_{\la_1,\sigma=\pm1} M_{\sigma} \,\Theta(\la_1\mu-\Ga_1)
            \left[\frac{ -\la_1}{ip_4-\la_1\Ga_1+\la_1\sigma\Ga_2}
                    +\frac{ \la_1}{ip_4+\la_1\Ga_1-\la_1\sigma\Ga_2}\right],
\nn\\[4pt]    \Delta_\mu \Sigma(p)&=&    -8e^2\int\frac{d^2q}{(2\pi)^2} \,
            \frac14\sum_{\la_1,\sigma=\pm1}    \frac{\sigma}{\Ga_1\Ga_2} \,
               \Theta(\la_1\mu-\Ga_1)
            \left[\frac{ -\la_1}{ip_4-\la_1\Ga_1+\la_1\sigma\Ga_2}
                    +\frac{ \la_1}{ip_4+\la_1\Ga_1-\la_1\sigma\Ga_2}\right].
\label{A.28}\eea
Now the expression in the square bracket is in fact independent on $\la_1$ which takes values $\pm1$ only. Thus the sum over $\la_1$ goes only over the step function and becomes
\be \sum_{\la_1=\pm1}\Theta(\la_1\mu-\Ga_1)=\Theta(\mu^2-\Ga_1^2)
\label{A.x}\ee
(again we used $\Ga_1>0$). We arrive at
\bea        \Delta_\mu \Pi_{00}(p)&=&-8e^2\int\frac{d^2q}{(2\pi)^2} \,\Theta(\mu^2-\Ga_1^2)
            \frac14\sum_{\la_1,\sigma=\pm1} M_{\sigma} \,
             \frac{ -\la_1} {ip_4-\la_1\Ga_1+\la_1\sigma\Ga_2} ,
\nn\\[4pt]    \Delta_\mu \Sigma(p)&=&    -8e^2\int\frac{d^2q}{(2\pi)^2} \,\Theta(\mu^2-\Ga_1^2)
            \frac14\sum_{\la_1,\sigma=\pm1}    \frac{\sigma}{\Ga_1\Ga_2} \,
             \frac{ -\la_1} {ip_4-\la_1\Ga_1+\la_1\sigma\Ga_2} .
\label{A.30}\eea
Now it is meaningful to turn to polar coordinates for the integration.  For this wee need to remove the square roots containing the cosine which can be done using
\be  \frac{ -\la_1} {ip_4-\la_1\Ga_1+\la_1\sigma\Ga_2}
=\frac{\Ga_1-i\la_1p_4+\sigma\Ga_2}{(ip_4-\la_1\Ga_1)^2-\Ga_2^2}.
\label{A.31}\ee
Carrying out the summation over $\sigma$ we get
\bea        \Delta_\mu \Pi_{00}(p)&=&-8e^2\int\frac{d^2q}{(2\pi)^2} \,\Theta(\mu^2-\Ga_1^2)
            \frac12\sum_{\la_1=\pm1}   \,
            \frac{1}{\Ga_1}\, \frac{Z} {N} ,
\nn\\[4pt]    \Delta_\mu \Sigma(p)&=&    -8e^2\int\frac{d^2q}{(2\pi)^2} \,\Theta(\mu^2-\Ga_1^2)
            \frac12\sum_{\la_1=\pm1}    \frac{1}{\Ga_1}\,
             \frac{1} {N}  ,
\label{A.33}\eea
with
\bea     Z&=&\Ga_1(\Ga_1-i\la_1p_4)+{\q}({\q}-{\p})+m^2,
\nn\\[4pt]   N&=&(ip_4-\la_1\Ga_1)^2-\Ga_2^2.
\label{A.34}\eea
The angular dependence is in ${\q}{\p}=\bar{q}\bar{p} \,\cos{\varphi}$ and in $\Ga_2^2=\Ga_1^2+\bar{p}^2-2\bar{q}\bar{p}\,\cos\phi$ and we rewrite
\bea        Z&=&\frac12\left(-N+(2\Ga_1-i\la_1p_4)^2-\bar{p}^2\right),
\nn\\[4pt]  N&=&-p_4^2-\bar{p}^2-2i\la_1p_4\Ga_1+2\bar{q}\bar{p}\,\cos\phi \equiv Q+a\,\cos\phi.
\label{A.35}\eea
Now the angular integration can be carried out by the formula
\be \int\frac{d\phi}{2\pi}\frac{1}{Q+a\,\cos\phi}=\frac{1}{\sqrt{Q^2-a^2}},
\label{A.36}\ee
and we get
\bea        \Delta_\mu \Pi_{00}(p)&=&\frac{-8e^2}{2\pi}\int_m^\mu d\Ga_1 \,
            \frac12\sum_{\la_1=\pm1}   \,
            \frac{1}{2}\left(-1+\frac{(2\Ga_1-i\la_1p_4)^2-\bar{p}^2}{ \sqrt{Q^2-a^2}}\right) ,
\nn\\[4pt]    \Delta_\mu \Sigma(p)&=& \frac{-8e^2}{2\pi}\int_m^\mu d\Ga_1 \,
            \frac12\sum_{\la_1=\pm1}   \,
             \frac{1} {\sqrt{Q^2-a^2}}  ,
\label{A.37}\eea
where $dq\,q=d\Ga_1\,\Ga_1$ was used.
The square root is uniquely defined at least for both real, $Q$ and $a$, with $|Q|>|a|$. For complex Q as in \Ref{A.35} the sign of the imaginary part of the square root follows the sign of the imaginary part of $Q$, i.e., it is $-\la_1$.

In fact we started the calculation of the polarization tensor in the Minkowskian region and turned to Euclidean momenta only temporarily. Now, after carrying out the $q_4$-integration we turn back by substituting $p_4=-ip_0$. Then $Q$ in \Ref{A.35} becomes
$ Q=p^2-2\la_1p_0\Ga_1$,
where \Ref{A.1b} was used. The  radicand in the denominators in \Ref{A.37} can be rewritten,
\be Q^2-a^2=
    p^2\bar{p}^2\left[\left(\frac{2\Ga_1-\la_1 p_0}{\bar{p}}\right)^2
    +\frac{4m^2-p^2}{p^2}\right]
\equiv p^2\bar{p}^2(x^2+x_0^2)
\label{A.38}\ee
with
\be x=\frac{2\Ga_1-\la_1p_0}{\bar{p}},\qquad x_0^2=\frac{4m^2}{p^2}-1.
\label{A.xx0}\ee
Under the rotation $p_4=-ip_0$, for $\la_1=+1$,  the $Q$ moves toward the  real axis.
For $|Q|>a$, which we assume in the following,  the square root $\sqrt{Q^2-a^2}$ becomes real and its sign follows the sign of $Q$. Thus we get
\be \sqrt{Q^2-a^2}\to {\rm sign}(Q)p\bar{p}\sqrt{x^2+x_0^2}
\label{A.39}\ee
and the last square root must be taken positive.
Rewriting in the first line in \Ref{A.37},
\be (2\Ga_1-i\la_1p_4)^2-\bar{p}^2
=\bar{p}^2\left[\left(\frac{2\Ga_1-\la_1 p_0}{\bar{p}}\right)^2-1\right]
=\bar{p}^2(x^2-1),
\label{A.40}\ee
we get from \Ref{A.37}
\bea        \Delta_\mu \Pi_{00}(p)&=&\frac{-8e^2}{2\pi}\int_m^\mu d\Ga_1 \,
            \frac12\sum_{\la_1=\pm1}   \,
            \frac{1}{2}\left(-1+{\rm sign}(Q) \frac{\bar{p}}{p}\frac{(x^2-1)}{ \sqrt{x^2+x_0^2}}\right) ,
\nn\\[4pt]    \Delta_\mu \Sigma(p)&=& \frac{-8e^2}{2\pi}\int_m^\mu d\Ga_1 \,
            \frac12\sum_{\la_1=\pm1}   \,
             \frac{{\rm sign}(Q)} {p\bar{p}\sqrt{x^2+x_0^2}} .
\label{A.41}\eea
Now the last integration can be carried out using the indefinite integrals
\bea \int\frac{dx}{\sqrt{x^2+x_0^2}}  &=&  {\rm arcsinh}\frac{x}{x_0},
\nn\\[4pt]  \int dx\frac{x^2-1}{\sqrt{x^2+x_0^2}}  &=&
    \frac12\left[x\sqrt{x^2+x_0^2}-(2+x_0^2) {\rm arcsinh}\frac{x}{x_0}\right]
\label{A.42}\eea
and $d\Ga_1=dx \,\bar{p}/{2}$.

Denoting the indefinite integrals by $\Pi_{00}(p;\Ga_1)$ and $\Sigma(p;\Ga_1)$ we get
\bea    \Pi_{00}(p;\Ga_1) &=&
            \frac{-8e^2\bar{p}}{8\pi}\frac12\sum_{\la_1=\pm1}
            \left[-x+{\rm sign}(Q)\frac{\bar{p}}{2p}\left(x\sqrt{x^2+x_0^2}
                        -(2+x_0^2){\rm arcsinh}\frac{x}{x_0}\right)\right],
\nn\\[4pt]  \Sigma(p;\Ga_1) &=&\frac{-8e^2}{4\pi p}\frac12\sum_{\la_1=\pm1}{\rm sign}(Q)\
                    {\rm arcsinh}\frac{x}{x_0}.
\label{A.43}\eea
Taken in the corresponding boundaries these give
%
\bea \Delta_\mu \Pi_{00}(p)&=& \Pi_{00}(p;\mu)-\Pi_{00}(p;m),
\nn\\[4pt]  \Delta_\mu \Sigma(p)&=& \Sigma(p;\mu)-\Sigma(p;m) .
\label{A.44}\eea
We will see that with the functions \Ref{A.43}  just that functions appear which were defined in \Ref{A.17}.

For this to show we consider these functions for   $\Ga_1= m$ and show that they coincide with the corresponding vacuum expressions in \Ref{A.25}.
In \Ref{A.43}, the dependence on $\Ga_1$ is in the variable $x$, defined in \Ref{A.38}, only. For $\Ga_1=m$ , which is equivalent to $q=0$ (see \Ref{A.Ga}), we get $x= (2m-\la_1 p_0)/{\bar{p}}$ and note
\be x^2+x_0^2 = \frac{Q^2}{p^2\bar{p}^2}.
\label{A.45}\ee
This follows with \Ref{A.Ga} from \Ref{A.35} for $q=0$. We have to take the square root,
\be {\rm sign}(Q)\sqrt{x^2+x_0^2}=\frac{ Q}{p\bar{p}} ,
\label{A.46}\ee
where the sign follows from \Ref{A.39}.
Further, with $\Ga=m$ and $p_4=-i\om$, we get from \Ref{A.35}
\be Q=p^2-2\la_1 p_0m,
\label{A.47}\ee
where $p^2=p_0^2-\bar{p}^2$ from \Ref{A.1b}. These formulas allow to express
\be
 \frac{ {\rm sign}(Q)\sqrt{x^2+x_0^2}}{x} =\frac{p^2-2\la_1p_0m}{p(2m-\la_1p_0)}.
\label{A.48}\ee
Further we use ${\rm arcsinh }z={\rm arccoth}\sqrt{z^{-2}+1}$ and ${\rm arccoth}\frac{a-b}{1-ab}={\rm arctanh}\,a-{\rm arccoth}\frac{1}{b}$ to get
\bea  {\rm arcsinh}\frac{x}{x_0} &=& {\rm arccoth}\frac{ \sqrt{x^2+x_0^2}}{x},
\nn\\[4pt]  &=& {\rm sign}(Q)\ {\rm arccoth}\frac{p^2-2\la_1p_0 m}{p(2m-\la_1 p_0)},
\nn\\[4pt]  &=& {\rm arctanh}\frac{p}{2m}-\la_1{\rm arctanh}\frac{p}{p_0}.
\label{A.acs}\eea
Using the above relations we rewrite \Ref{A.43},
\bea    \Pi_{00}(p;m) &=&
            \frac{-8e^2\bar{p}}{8\pi}\frac12\sum_{\la_1=\pm1}
            \left[-\frac{2m-\la_1 p_0}{\bar{p}}
            \left(1-\frac{\bar{p}}{2p}\frac{p^2-2\la_1p_0 m}{p\bar{p}}\right)
            -\frac{ \bar{p}}{2p}\left(1+\frac{4m^2}{p^2}\right)
                \left(  {\rm arctanh}\frac{p}{2m}-\la_1{\rm arctanh}\frac{p}{p_0}\right)
            \right]
            ,
\nn\\[4pt]  \Sigma(p;m) &=&\frac{-8e^2}{4\pi p}\frac12\sum_{\la_1=\pm1}
                   \left( {\rm arctanh}\frac{p}{2m}- \la_1
                   {\rm arctanh}\frac{p}{p_0}\right).
\label{A.49b}\eea
The summation simplifies these expressions,
\bea    \Pi_{00}(p;m) &=&
            \frac{-8e^2\bar{p}^2}{16\pi p^3}
            \left( 2m p  -(p^2+4m^2)\,{\rm arctanh}\frac{p}{2m} \right)
            ,
\nn\\[4pt]  \Sigma(p;m) &=&\frac{-8e^2}{4\pi p }\, {\rm arctanh}\frac{p}{2m} .
\label{A.50}\eea
Now the statement is that Eqs.  \Ref{A.50} give just the vacuum contributions in Eq. \Ref{A.25}.
For  both expression this can be verified easily, especially for the scalar contribution. For the polarization tensor we mention Eqs.  (6) and (7) in \II~ and Eq. (A.16) in \cite{pyat09-21-025506} (with $\v=1$).


In this way, when adding \Ref{A.43} with the vacuum expression in \Ref{A.25}, the latter cancel and we are left with the  final representations
\bea
 \Pi_{00}(p;\mu) &=&\left\{
 {
            \frac{8e^2 }{4\pi  }
            \left[\mu-\frac12\sum_{\la_1=\pm1}\frac{{\rm sign}(Q)\, \bar{p}^2}{4p}
             \left(x\sqrt{x^2+x_0^2}-(2+x_0^2) \,{\rm arcsinh}\frac{x}{x_0} \right)\right]\quad (\mu>m),
  \atop \frac{-8e^2\bar{p}^2}{16\pi p^3}
            \left( 2m p  -(p^2+4m^2)\,{\rm arctanh}\frac{p}{2m} \right)\quad (\mu\le m),
  }\right.
\label{A.51a} \eea
and
\bea  \Sigma(p;\mu) &=&
 \left\{  {
            \frac{- 8e^2}{4\pi p }\, \frac12\sum_{\la_1=\pm1} {\rm sign}(Q)\ {\rm arcsinh}\frac{x}{x_0}\quad (\mu>m),
 \atop      \frac{-8e^2}{4\pi p }\, {\rm arctanh}\frac{p}{2m}\quad (\mu\le m),
 }\right.
\label{A.51}\eea
where now
\be x=\frac{2\mu-\la_1 p_0}{\bar{p}},\quad x_0^2=\frac{4m^2}{p^2}-1,\quad Q=p^2+2\la_1 \om \mu.
\label{A.52}\ee
It remains to calculate the tadpole contribution in \Ref{A.17}. We rewrite it for $T=0$,
\be  \Sigma_{\rm tp}(p;\mu)
        =-8e^2\int\frac{d q_4}{2\pi} \int\frac{d^2q}{(2\pi)^2}
        \frac{1}{ (iq_4-\mu +\Gamma_1)(iq_4-\mu -\Gamma_1)},
\label{A.53}\ee
where we wrote the denominator as a product. Now we apply Eq. \Ref{A.19},
\be  \Sigma_{\rm tp}(p;\mu)
        =-8e^2 \int\frac{d^2q}{(2\pi)^2}  \frac12
        \frac{{\rm sgn}(-\mu+\Ga_1)- {\rm sgn}(-\mu-\Ga_1)}{ -2\Ga_1}.
\label{A.54}\ee
The sign factors can be rewritten in the form
\be {\rm sgn}(-\mu+\Ga_1)- {\rm sgn}(-\mu-\Ga_1)=2-2\Theta(\mu^2-\Ga_1^2),
\label{A.55}\ee
allowing to split into the vacuum part,
\be  \Sigma_{\rm tp}^{(0)}(p)=
-8e^2\int\frac{d^{3-2\ep}q}{(2\pi)^{3-2\ep}}\,\frac{1}{-q_4^2-\bar{q}^2-m^2},
\label{A.56}\ee
and the addendum
\be \Delta_\mu\Sigma_{\rm tp}(p)=\frac{-8e^2}{2}\int\frac{d^2q}{(2\pi)^2}.
                \frac{\Theta(\mu^2-\Ga_1^2)}{\Ga_1}
\label{A.57}\ee
In the vacuum part we introduced dimensional regularization since the integral is divergent. Direct integration gives
\be  \Sigma_{\rm tp}^{(0)}(p)=
8e^2\frac{\Gamma(\ep-1/2)}{(4\pi)^{3/2-\ep}}=\frac{-8e^2}{4\pi}m+O(\ep).
\label{A.58}\ee
As usual, in even space dimension the dimensional regularization does not give a pole contribution.

In \Ref{A.57}, there is no angular dependence in the integrand and the $q$-integration is trivial,
\be \Delta_\mu\Sigma_{\rm tp}(p)=\frac{-8e^2}{4\pi}\,(\mu-m).
\label{A.59}\ee
Putting \Ref{A.58} and \Ref{A.59} together we get
\be \Sigma_{\rm tp}=\frac{-8e^2}{4\pi}\,\max(\mu,m),
\label{A.60}\ee
which is the final result for the tadpole contribution.

Now, from \Ref{A.12} with \Ref{A.58} and \Ref{A.57} we get for the trace of the polarization tensor
\bea \Pi_{\rm tr}(p;\mu) &=&
    \left\{ {        \frac{8e^2}{4\pi}
            \left(\mu-\frac{p^2+4m^2}{2p}\frac12\sum_{\la_1=\pm1}{\rm sign}(Q)\,
            {\rm arcsinh}\frac{x}{x_0}\right),\quad (\mu> m),
    \atop   \frac{ 8e^2}{8\pi p}
            \left( 2m p  -(p^2+4m^2)\,{\rm arctanh}\frac{p}{2m} \right),\quad (\mu\le m).
    }   \right.
\label{A.61}\eea
By equations \Ref{A.51} and \Ref{A.61} we have explicit expressions for the two form factors of the polarization tensor. In the course of the derivation, after Eq. \Ref{A.37}, we turned from the Euclidean to the Minkowski region. Thereby, especially in Eq. \Ref{A.42}, we assumed $x_0$ to be real. With \Ref{A.xx0} this implies $p^2<(2m)^2$, i.e., the region below the threshold, which is, in terms of frequency at $p_0=\om_s$ with $\om_s\equiv\sqrt{\bar{p}^2+4m^2}$.
Thus, Eqs. \Ref{A.51} and \Ref{A.61}, as written, are in this region in terms of real functions.

The above formulas allow for an easy analytic continuation to the region above threshold, i.e., to $p_0>\om_s$. In the complex $p_0$-plane, the polarization tensor has a cut starting from $\om_s$. We consider the continuation in the upper half plane, i.e., for $\Im p_0>0$. Thereby we have
\be x_0=\sqrt{\frac{(2m)^2}{\p_0^2-\bar{p}^2}-1} \to -i y_0\quad\mbox{with}\quad y_0=\sqrt{\frac{\p_0^2-(\bar{p}^2+(2m)^2)}{\p_0^2-\bar{p}^2}}.
\label{A.62}\ee
Note $p_o^2>\bar{p}^2$ because we are in the Minkowskian region.  For the continuation in Eqs.  \Ref{A.51} and \Ref{A.61} we use the relations
\bea {\rm arcsinh}(iz) &=&={\rm sign}(z)\left(i\frac{\pi}{2}
                                    +{\rm arccosh}(|z|)\right),
\nn\\   {\rm arctanh}(iz)&=&i\frac{\pi}{2}+{\rm arctanh}\left(\frac{1}{z}\right)
\label{A.63}\eea
$(|z|>1) $, where we have to insert $z\to\frac{x}{y_0}$. Using \Ref{A.39} we note $x^2-y_0^2=(Q^2-a^2)/(p^2\bar{p}^2)$, thus $|x/y_0|>1$.
Using Eqs.  \Ref{A.52} and \Ref{A.35} (for $a$), $Q^2-a^2$ can be seen to be a fourth order polynom in $p_0$. Its roots are
\be \om_{\sigma'}^{\sigma}(\la)=\la\mu+\sigma'\sqrt{\mu^2+\bar{p}^2+2\sigma\bar{p}\,k_F}
    \ \ (\sigma,\sigma'=\pm1)
\label{A.63a}\ee
Analyzing these roots, the restriction $p_0<\om_+^-$ follows, together with $\bar{p}<p_0$. Further we need the sign of $Q$, Eq. \Ref{A.52}. It is a second order polynom in $p_0$ and from analyzing its roots under the above restrictions, ${\rm sign}(Q)=\la_1$ follows.

Using now \Ref{A.63} in the upper lines of Eqs.  \Ref{A.51} and \Ref{A.61}, and accounting for the sum over $\la_1$,  the imaginary contributions cancel. As a result we get
\be \sum_{\la_1}{\rm sign}(Q)\,{\rm arcsinh}\left(\frac{x}{x_0}\right)\to \sum_{\la_1}\la_1{\rm arccosh}\left(\left|\frac{x}{y_0}\right|\right)
\label{A.64}\ee
for the transition to above the threshold. In all other places in \Ref{A.51} and \Ref{A.61} we substitute simply $x_0^2\to -y_0^2$.

In the expressions for $m>\mu$, i.e., in the second lines of Eqs.  \Ref{A.51} and \Ref{A.61}, we use the second line in \Ref{A.63}. Together, the final formulas for the from factors of the polarization tensor above the threshold read
\bea
 \Pi_{00}(p;\mu) &=&\left\{
 {
            \frac{8e^2 }{4\pi  }
            \left[\mu+\frac{\bar{p}^2}{4p}\frac12\sum_{\la_1=\pm1}\la_1
             \left(x\sqrt{x^2-y_0^2}-(2-y_0^2) \,{\rm arccosh}\left(\left|\frac{x}{y_0} \right|\right) \right)\right]\quad (\mu>m),
  \atop \frac{-8e^2\bar{p}^2}{16\pi p^3}
            \left( 2m p  -(p^2+4m^2)\,\left(i\frac{\pi}{2}+{\rm arctanh}\frac{2m}{p}\right) \right)\quad (\mu\le m),
  }\right.
\label{A.65}\eea
and
\bea \Pi_{\rm tr}(p;\mu) &=&
    \left\{ {        \frac{8e^2}{4\pi}
            \left(\mu+\frac{p^2+4m^2}{p}\frac12\sum_{\la_1=\pm1}\la_1\,
            {\rm arccosh}\left(\left|\frac{x}{y_0} \right|\right)\right),\quad (\mu> m),
    \atop   \frac{ 8e^2}{8\pi p}
            \left(  2m p  -(p^2+4m^2)\,\left(i\frac{\pi}{2}+{\rm arctanh}\frac{2m}{p}\right)  \right),\quad (\mu\le m).
    }   \right.
\label{A.66}\eea
The upper lines of these equations are real for
\be \sqrt{\bar{p}^2+(2m)^2}<p_0<\om_1^-\,.
\label{A.67}\ee
This defines the region where the polarization tensor is real which is of relevance for the plasmons. In all other regions, one has either an imaginary part or, for physical values of $\al$ and $v$, the frequency is outside the validity region of the Dirac model.

As for the lower lines, i.e., for $m>\mu$ which is equivalent to no chemical potential, there is the usual imaginary part above threshold resulting from particle creation.
%
\section{The case $m=0$}
Her we specify the formulas \Ref{A.51a} and \Ref{A.61} for the case $m=0$. In this case only the upper lines of these formulas apply.
For the variable $x$ still \Ref{A.52} applies. For $x_0$ we get $x_0^2=-1$ and  taking $\om$ with positive imaginary part, $x_0=-i$. Further we need
\be {\rm arcsinh}(ix)=\left\{ {i \arcsin(x), \ ~~~~~~~~~~~~~~~~~~~~~~ \mbox{for}\ |x|<1,\atop
                            {\rm sign}(x)\left(\frac{i\pi}{2}+{\rm arccosh}(|x|)\right),
                             \ \mbox{for}\ |x|>1}.
                            \right.
\label{B.1}\ee
For $|x|<1$ we observe imaginary parts from this formula and from the square root in \Ref{A.51a} as well and in this region we cannot expect a stable solution. In opposite, for $|x|>1$, the expressions become real after the sum over $\la_1$ is done, which makes the contribution from $i\pi/2$ vanishing.  Further, if we consider $\mu>0$ only, we have $x>1$ and can drop the signs in \Ref{B.1}. In this way, in the massless case the polarization tensors become
\bea        \Pi_{00}(p;\mu)&=&\frac{2e^2}{\pi}\left[\mu-\frac{\bar{p}^2}{4p}\frac12\sum_{\la_1=\pm1}
{\rm sign}(Q)\, \left(x\sqrt{x^2-1}-{\rm arccosh}(x)\right)\right],
\nn\\   \Pi_{\rm tr}(p;\mu)&=&\frac{2e^2}{\pi}\left[\mu-\frac{p}{2}\frac12\sum_{\la_1=\pm1}
{\rm sign}(Q)\,  \,{\rm arccosh}(x) \right],
\label{B.2}\eea
with
\be     x=\frac{2\mu-\la_1p_0}{\bar{p}}
\label{B.3}\ee
which is the same as \Ref{A.52}.

\bibliography{C:/Users/bordag/WORK/Literatur/bib/papers,C:/Users/bordag/WORK/Literatur/Bordag,C:/Users/bordag/WORK/Literatur/bib/libri}
\end{document}